\documentclass[preprint,12pt]{elsarticle}
\usepackage{graphicx}
\usepackage{amssymb}
\journal{Elsevier}

\begin{document}

\begin{frontmatter}

\title{Quantum Correlations in non-Markovian Environments}

\author{Ferdi Altintas\corref{cor1}}\ead{altintas$_{\_\ }$f@ibu.edu.tr}\cortext[cor1]{Corresponding author. Tel.:$+$90 544 5460504; fax:$+$90 374 2534642.}
\author{Resul Eryigit}
\address{Department of Physics, Abant Izzet Baysal University, Bolu, 14280-Turkey.}
\begin{abstract}
We have studied the analytical Markovian and non-Markovian dynamics of quantum correlations, such as entanglement, quantum discord and Bell nonlocalities for three noisy qubits. Quantum correlation as measured by quantum discord is found to be immune to death contrary to entanglement and Bell nonlocality for initial GHZ- or W-type mixed states.
\end{abstract}
\end{frontmatter}

\section{Introduction}

Entanglement, quantum discord and Bell nonlocalities and their dynamics under the influence of external noise have been an active area of research, recently~\cite{bb1,bb2,bb4,bb3,lsz,bb5,bb6,purity,projectors,bb7,bb8,jql,ferraro}. Although  entanglement was considered the most important form of quantum correlations for carrying out quantum computations~\cite{ue1} or other quantum tasks, such as quantum teleportation~\cite{ue2} and quantum cryptography~\cite{ue3}, it was shown by Datta~{\it et al.}  that states that have no entanglement but non-zero quantum discord still can be used to perform useful quantum tasks~\cite{spc}.

Entanglement, quantum discord and Bell nonlocality of a given state are all different aspects of quantum correlations in that state. Entanglement refers to  separability of the states while quantum discord is the difference between the total and the classical correlations for a given state as measured by von-Neumann entropy. Quantum discord is an independent measure of nonclassical correlations which might include entanglement, but the relation between entanglement and quantum discord is complicated for a general states~\cite{qd,qd2,qd3,qd4}. Bell nonlocality is involved with the violations of Bell inequalities which identify the entangled mixed states whose correlations can be reproduced by a classical local model~\cite{bb9}.

All quantum systems have an environment that entangles with the system and drives it to an incoherent state. It was shown by  Yu and  Eberly that while coherence of a single qubit goes to zero exponentially, the entanglement between two such qubits can cease to exist in a finite time, a phenomenon called entanglement sudden death~(ESD)~\cite{esd1}. After Ref.~\cite{esd1}, a number of studies on the dynamics of other types of quantum correlations, such as quantum discord and Bell nonlocalities under Markovian as well as non-Markovian memory types were carried out to investigate whether sudden death phenomena exist for those correlations~\cite{bb1,bb2,bb5,purity,projectors,bb7,bb8,jql,ferraro}.  Wang~{\it et al.} have considered the effects of a non-Markovian dissipative environment on the dynamics of a two qubit system for Bell-like or extended Werner-like initial states and found that quantum discord is more robust compared to the entanglement as measured by concurrence~\cite{bb7}. Werlang~{\it et al.} studied the dynamics of quantum discord and entanglement for the Markovian environment with depolarizing, dephasing and generalized amplitude damping channels and found that whenever there is ESD, the discord decreases only exponentially~\cite{bb8}. A number of work dealt with the comparision of the behavior of entanglement and Bell nonlocality for two qubits in various environments~\cite{bb5,bb6,purity,jql,bb9}.  Werner has demonstrated that mixed-entangled states might not violate Bell inequalities~\cite{bb9}. Bellomo~{\it et al.} have found that Bell inequality might not be violated for a state with high values of entanglement for a two qubit system subject to amplitude damping~\cite{bb5}. 

Although the relation between the memory properties (Markovian and non-Markovian) and type of correlation and the resulting dynamics is investigated for a bipartite systems by a large number of  groups, the works on multipartite systems are scarce~\cite{bb1,bb2,bb4,bb3,lsz}. Among them, Ann and Jaeger examined the finite time destruction of multipartite Bell inequality violation in three-qubit systems initially prepared in GHZ and W pure states under the multi-local asymptotic dephasing noise in Markovian approximation~\cite{bb1,bb2}. Moreover, Qui~{\it et al.} showed that the tripartite Bell inequality violations can be fully destroyed in a finite time for a general pure state in a nonlocal antiferromagnetic environment under an external magnetic field by using spin wave approximation~\cite{bb4}. Also, Yang~{\it et al.} demonstrated the three qubit Bell nonlocality sudden death in Tavis-Cummings model with cavity loss for W-like initial state by solving the Schr\"{o}dinger equation exactly~\cite{bb3}. Liu, Shao and Zou considered the dynamics of Bell-nonlocality for a three-qubit Heisenberg XY chain in the presence of an external magnetic field for pure GHZ-state and found that the multipartite Bell-inequality violations can be destroyed in a finite time because of intrinsic decoherence~\cite{lsz}. However, none of them compares the dynamics of quantum correlations.

In the present study, we have analyzed the dynamics of Bell nonlocalities  as measured by Mermin-Ardehali-Belinksii-Klyshko~(MABK) and Svetlichny inequalities\cite{bbell1,bbell2,bbell3,bbell4}, quantum discord and entanglement as measured by concurrence for bipartitions and tripartite negativity for tripartite states for a system of three qubits which have energy levels that are stochastic with Ornstein-Uhlenbeck type correlations. The analytic expressions for quantum discord, Bell nonlocalities as well as concurrence and tripartite negativity are derived for W- and GHZ-type initial states by exploiting a procedure based on the knowledge of single-qubit dynamics~\cite{gprocedure}. We consider both Markovian and non-Markovian time evolution and compare and contrast the character of time dependence of quantum correlations.

The organization of this paper is as follows. In Sec.~\ref{model}, we introduce the model and its solution using the procedure analyzed in Appendix~\ref{procedure}. In Sec.~\ref{qcorrelation}, we show explicit analytic calculations of quantum discord, concurrence, tripartite negativity and Bell nonlocalities for three-qubit system initially prepared in GHZ- and W-type states and explore the effects of non-Markovianity and mixedness on quantum correlations. We conclude with a summary of important findings in Sec.~\ref{conc}.

\section{The Model and its Solution}
\label{model}

The model we consider in this paper is three uncoupled qubits interacting with their noisy environments independently which cause time-dependent fluctuations in their energy levels. The Hamiltonian for this model is given by~\cite{noiseonly} (we set $\hbar=1$):
\begin{eqnarray}
\label{hamiltonian2}
\hat{H}(t)=\frac{\Omega_A(t)}{2} \hat{\sigma}_z^A+\frac{\Omega_B(t)}{2} \hat{\sigma}_z^B+\frac{\Omega_C(t)}{2} \hat{\sigma}_z^C,
\end{eqnarray}
where $\hat{\sigma}_z$ is the usual Pauli spin operator in z-direction, $\Omega_{A,B,C}(t)$ are the independent fluctuations of the transition frequencies obeying  non-Markovian approximation with mean value properties 
\begin{eqnarray}
\label{mean}
M\{\Omega_i(t)\}=0,\\
M\{\Omega_i(t)\Omega_i(s)\}&=&\alpha(t-s)\nonumber\\
&=&\frac{\Gamma_i \gamma}{2} e^{-\gamma \left|t-s\right|} , i=A,B,C,
\end{eqnarray}
where $M\{ ...\}$ stands for the statistical mean over the noises $\Omega_A(t)$, $\Omega_B(t)$ and $\Omega_C(t)$. Here $\Gamma_i (i=A,B,C)$ are the damping rates due to the coupling to the environments, $\gamma$ is the noise bandwidth which determines the finite correlation time of the environmental noise $(\tau_c=\gamma^{-1})$ and $\alpha(t-s)$ is the reservoir correlation function. For simplicity, we will take the noise properties to be the same for $A$, $B$ and $C$ (e.g., $\Gamma_A=\Gamma_B=\Gamma_C\equiv\Gamma$). And note that in the limit $\gamma\rightarrow\infty  (\tau_c\rightarrow0)$, Ornstein-Uhlenbeck noise reduces to the well-known Markovian case~\cite{noiseonly2}:
\begin{eqnarray}
\label{markov}
\alpha(t-s)=\Gamma\delta(t-s).
\end{eqnarray}
For the total system described by the Hamiltonian~(\ref{hamiltonian2}), the stochastic Schr\"{o}dinger equation is given by 
\begin{eqnarray}
\label{schrödinger}
i{d \over dt}\left|\Psi(t)\right\rangle=\hat{H}(t)\left|\Psi(t)\right\rangle,
\end{eqnarray}
with solution 
\begin{eqnarray}
\label{solution}
\left|\Psi(t)\right\rangle=\hat{U}(t,\Omega_A,\Omega_B,\Omega_C)\left|\Psi(0)\right\rangle,
\end{eqnarray}
where the stochastic propagator $\hat{U}(t,\Omega_A,\Omega_B, \Omega_C)$ is given by
\begin{eqnarray}
\label{propagator}
\hat{U}(t,\Omega_A,\Omega_B,\Omega_C )=e^{-\frac{i}{2}\int_0^t(\Omega_A(s)\hat{\sigma}_z^A+\Omega_B(s)\hat{\sigma}_z^B+\Omega_C(s)\hat{\sigma}_z^C)ds}.
\end{eqnarray}
The reduced density matrix for spins $A$, $B$ and $C$ is then obtained from the statistical mean
\begin{eqnarray}
\label{reduceddensity}
\hat{\rho}(t)=M\{ \left|\Psi(t)\right\rangle\left\langle \Psi(t)\right| \}.
\end{eqnarray}

Since we assume independent reservoirs and the qubits interact only with their own independent reservoir, we can use the procedure developed by Belloma {\it et al.} to obtain the time dependent density matrix of three qubit system from the knowledge of the dynamics of single qubit in an environment~\cite{gprocedure}. Using the stochastic Schr\"{o}dinger equation~(\ref{schrödinger}), the  master equation for the single qubit reduced density matrix can be derived as~\cite{master1,master2,master3} :
\begin{eqnarray}
\label{mastereqn}
{d \over dt}\hat{\rho}(t)=-\frac{G(t)}{2}(\hat{\rho}(t)-\hat{\sigma}_z\hat{\rho}(t)\hat{\sigma}_z),
\end{eqnarray}
where
\begin{eqnarray}
\label{memoryinformation}
G(t)=\int_0^t\alpha(t-s)ds= \frac{\Gamma}{2}(1-e^{-\gamma t}).
\end{eqnarray}
Here $G(t)$ is a time-dependent coefficient which includes the memory information of the environmental noise. The analytic solution of Eq.~(\ref{mastereqn}) and the time-dependent density matrix for tripartite system obtained according to procedure of Ref.~\cite{gprocedure} is outlined in Appendix~\ref{procedure}.

Since we assume the noise properties to be the same for all qubits, the parameters in Eq.~A(\ref{sqelements}) can be obtained from the solution of Eq.~(\ref{mastereqn}) with the help of Eqs.~A(\ref{sqdms}) and~A(\ref{sqelements}) as:
\begin{eqnarray}
\label{sqparameters}
u_t^A&=&u_t^B=u_t^C=1,\nonumber\\
v_t^A&=&v_t^B=v_t^C=0,\nonumber\\
z_t^A&=&z_t^B=z_t^C=e^{-f(t)},
\end{eqnarray}
where
\begin{eqnarray}
f(t)&=&\int_0^tG(s)ds\nonumber\\
&=&\frac{\Gamma}{2}\left(t+\frac{1}{\gamma}(e^{-\gamma t}-1)\right).
\end{eqnarray}
Using Eqs.~A(\ref{tqdelements}),~A(\ref{tqoelements}) and ~(\ref{sqparameters}), the matrix elements of the reduced density matrix for three-qubit can be determined:
\begin{eqnarray}
\label{tqrdme}
\rho_{ii}(t)&=&\rho_{ii}(0),\qquad  \{i=1,2,\ldots,8\}, \nonumber\\
\rho_{ij}(t)&=&\rho_{ij}(0)e^{-f(t)},\quad \{ij=12,13,15,24,26,34,37,48,56,57,68,78\}, \nonumber\\
\rho_{ij}(t)&=&\rho_{ij}(0)e^{-2f(t)},\quad \{ij=14,16,17,23,25,28,35,38,46,47,58,67\}, \nonumber\\
\rho_{ij}(t)&=&\rho_{ij}(0)e^{-3f(t)},\quad \{ij=18,27,36,45\}.
\end{eqnarray}

\section{Correlation Measures}
\label{qcorrelation}
In this section, we briefly review the correlation measures considered in the present study.

{\bf Quantum Entanglement:} For a pair of qubits, the concurrence as a measure of entanglement is well-defined. It is introduced by Wootters~\cite{wooters} and defined as
\begin{eqnarray}
\label{con1}
C=\max\{0,\sqrt{\lambda_1}-\sqrt{\lambda_2}-\sqrt{\lambda_3}-\sqrt{\lambda_4}\},
\end{eqnarray}
where the quantities $\lambda_i(i=1,2,3,4)$ are the eigenvalues of the matrix
\begin{eqnarray}
\hat{R}=\hat{\rho}_{AB}(\hat{\sigma}_y^A\otimes\hat{\sigma}_y^B)\hat{\rho}_{AB}^*(\hat{\sigma}_y^A\otimes\hat{\sigma}_y^B),
\end{eqnarray}
in descending order and $\hat{\rho}_{AB}^*$ is the conjugate of $\hat{\rho}_{AB}$. For the density matrix with X-form~\cite{noiseonly},
\begin{eqnarray}
\label{xmatrix}
\hat{\rho}_{AB}=\left [ \begin{array}{cccc}
  a & 0 & 0  & w \\
  0  & b & z  & 0 \\ 0  & z & c  & 0 \\ w  & 0 & 0  & d
\end{array} \right] \ ,
\end{eqnarray}
the concurrence function~(\ref{con1}) has a simple analytic form:
\begin{eqnarray}
\label{con2}
C=2\max\{0,z-\sqrt{a d},w-\sqrt{b c} \}.
\end{eqnarray}

On the other hand, for the three-qubit case concurrence is not well-defined. Its calculation is based on numerical optimization procedure which does not guarantee exact results~\cite{minter,sf}. Instead, for mixed states of three qubit systems, the tripartite negativity $N$ was introduced by Sabin and Garcia-Alcaine~\cite{sga} as
\begin{eqnarray}\label{tqnega}
N=\left(N_{A-BC}N_{B-AC}N_{C-AB}\right)^{1/3},
\end{eqnarray}
where $N_{I-JK}$ with $I=A,B,C$ and $JK=BC,AC,AB$ is the negativity with respect to the subsystem $I$ and defined as~\cite{nega}
\begin{eqnarray}\label{opnega}
N_{I-JK}=-\displaystyle\sum_i\sigma_i(\hat{\rho}^{tI}),
\end{eqnarray}
where $\sigma_i(\hat{\rho}^{tI})$ are the negative eigenvalues of the partial transpose $\hat{\rho}^{tI}$ of the system density matrix with respect to the subsystem $I$~\cite{partial}.

One should note that for the mixed states, $N$ is not able to quantify multipartite entanglement fully, but its positivity ensures that the state under consideration is not separable~\cite{clp}. $N$ is used to study  the time evolution of entanglement for a number of tripartite systems~\cite{clp,enega,weinstein1,weinstein2}. For the type of GHZ and W states and the dynamics considered in the present work  $N$ is the same as $N_{I-JK}$~\cite{weinstein2}.

{\bf Quantum Discord:} For two-qubit systems, quantum discord as a measure of quantum correlation was introduced by Ollivier and Zurek~\cite{qd}. It is defined as the difference between two expressions of mutual information: total and classical correlations, namely,
\begin{eqnarray}
\label{qdiscord}
D=I(\hat{\rho}_{AB})-J(\hat{\rho}_{AB}).
\end{eqnarray}
Here $I(\hat{\rho}_{AB})$ is the total correlation between two subsystems defined as
\begin{eqnarray}
\label{tdiscord}
I(\hat{\rho}_{AB})=S(\hat{\rho}_A)+S(\hat{\rho}_B)-S(\hat{\rho}_{AB}),
\end{eqnarray}
where $S(\hat{\rho})=-Tr(\hat{\rho}\log_2\hat{\rho})$ is the von Neumann entropy and $\hat{\rho}_A(\hat{\rho}_B)$ is the reduced density matrix of $\hat{\rho}_{AB}$ obtained by tracing out $B(A)$~\cite{partial}. The other quantity $J(\hat{\rho}_{AB})$ is the classical correlation between $A$ and $B$ as the maximum information one can get from $A$ by measuring $B$. It is defined as 
\begin{eqnarray}
\label{cdiscord}
J(\hat{\rho}_{AB})=\displaystyle\max_{\{\hat{\Pi}_i\}}\{S(\hat{\rho}_A)-\displaystyle\sum_i p_i S(\hat{\rho}_{A|_i})\},
\end{eqnarray}
where $\hat{\rho}_{A|_i}=\frac{Tr_B(\hat{\Pi}_i\hat{\rho}_{AB}\hat{\Pi}_i)}{Tr_{AB}(\hat{\Pi}_i\hat{\rho}_{AB}\hat{\Pi}_i)}$ and $\hat{\Pi}_i$ is a set of projectors defined as $\hat{\Pi}_i=\hat{I}\otimes\left|i\right\rangle\left\langle i\right|~(i=1,2)$ where $\left|1\right\rangle=\cos\theta\left|+_z\right\rangle+e^{i\phi}\sin\theta\left|-_z\right\rangle$ and $\left|2\right\rangle=\sin\theta\left|+_z\right\rangle-e^{i\phi}\cos\theta\left|-_z\right\rangle$ are the orthogonal states~\cite{projectors}. The projector operators are used to measure subsystem $B$, corresponding to the outcome {\it i} with probability $p_i=Tr_{AB}(\hat{\Pi}_i\hat{\rho}_{AB}\hat{\Pi}_i)$.

In general, it is hard to calculate the analytic expression for the quantum discord. However, if the reduced density matrix of two qubits has X-form as Eq.~(\ref{xmatrix}) with $b=c$, the quantum discord has a simple analytic form~\cite{projectors}:
\begin{eqnarray}
\label{qdiscord2}
D=\min\{D_1,D_2\},
\end{eqnarray}
where 
\begin{eqnarray}
D_1&=&S(\hat{\rho}_A)-S(\hat{\rho}_{AB})-a\log_2\left(\frac{a}{a+b}\right)-b\log_2\left(\frac{b}{a+b}\right)\nonumber\\
&-&d\log_2\left(\frac{d}{b+d}\right)-b\log_2\left(\frac{b}{d+b}\right),\nonumber\\
D_2&=&S(\hat{\rho}_A)-S(\hat{\rho}_{AB})-\frac{1}{2}(1+\kappa)\log_2\left(\frac{1}{2}(1+\kappa)\right)\nonumber\\
&-&\frac{1}{2}(1-\kappa)\log_2\left(\frac{1}{2}(1-\kappa)\right),
\end{eqnarray}
where $\kappa^2=(a-d)^2+4(|z|+|w|)^2$.

{\bf  Bell-Nonlocality:} For three-qubit case, violation of Bell nonlocality can be used as a measure of quantum correlations. From this point,  we choose two kinds of nonlocality measures. The first one is introduced by Mermin-Ardehali-Belinksii-Klyshako (MABK)~\cite{bb1,bb2,bbell1,bbell2,bbell3}. It is easily computable for three-qubit case and violated whenever $\left|\left\langle \hat{B}\right\rangle_{\rho}\right|=\left|Tr(\hat{B}\hat{\rho}(t))\right|>1$ where MABK operator is given by
\begin{eqnarray}
\label{mabk}
\hat{B}=\frac{1}{2}\left(\hat{M}_A\hat{M}_B\hat{M}_C'+\hat{M}_A\hat{M}_B'\hat{M}_C+\hat{M}_A'\hat{M}_B\hat{M}_C-\hat{M}_A'\hat{M}_B'\hat{M}_C'\right).
\end{eqnarray}
The second inequality is put forward by Svetlichny~\cite{bb1,bb2,bbell4} and denotes genuine tripartite Bell nonlocality if $\left|\left\langle \hat{S}\right\rangle_{\rho}\right|=\left|Tr(\hat{S}\hat{\rho}(t))\right|>4$ where the Svetlichny operator is
\begin{eqnarray}
\label{sve}
\hat{S}&=&\hat{M}_A\hat{M}_B\hat{M}_C+\hat{M}_A\hat{M}_B\hat{M}_C'+\hat{M}_A\hat{M}_B'\hat{M}_C+\hat{M}_A'\hat{M}_B\hat{M}_C\nonumber\\
&-&\hat{M}_A'\hat{M}_B'\hat{M}_C'-\hat{M}_A'\hat{M}_B'\hat{M}_C-\hat{M}_A'\hat{M}_B\hat{M}_C'-\hat{M}_A\hat{M}_B'\hat{M}_C'.
\end{eqnarray}
where $\hat{M}_K$ is the measurement operator for the $K^{th}$ qubit, the primed and unprimed terms correspond to different measurement directions for the measuring party. The measurement operator for each successive subsystem is obtained from the preceding one by a rotation: 
\begin{eqnarray}
\left(\begin{array}{c}
\hat{M}_K \\ \hat{M}_K'\end{array}\right)=\left(\begin{array}{cc}
\cos\theta_K & -\sin\theta_K\\
\sin\theta_K & \cos\theta_K\end{array}\right) \left(\begin{array}{c}
\hat{M}_A \\ \hat{M}_A'\end{array}\right),
\end{eqnarray}
where $\theta_K~(K=B,C)$ are the rotation angles~\cite{bb1,bb2}.

To demonstrate the finite time loss of Bell nonlocality in both Markovian and non-Markovian environments, we shall assume two different nonlocality regimes satisfying:
\begin{eqnarray*}
(i)\quad \left|\left\langle \hat{B}\right\rangle_{\rho}\right|>1,\quad \left|\left\langle \hat{S}\right\rangle_{\rho}\right|>4,\\
(ii)\quad \left|\left\langle \hat{B}\right\rangle_{\rho}\right|\leq1,\quad \left|\left\langle \hat{S}\right\rangle_{\rho}\right|\leq4,
\end{eqnarray*}
where the regime~(i) shows that the state $\hat{\rho}(t)$ is genuinely tripartite Bell nonlocal. Also we choose the rotation angles in which $\left|\left\langle \hat{B}\right\rangle_{\rho}\right|$ and $ \left|\left\langle \hat{S}\right\rangle_{\rho}\right|$ have maximum values at $t=0$.

In the following, we will consider the dynamics of tripartite GHZ- and W-type states. As is well known, GHZ-state cannot be transformed into W-state by local operations and classical communication~\cite{wg}. This means they bear incompatible multipartite correlations. From this point of view, one may expect that the two initial states might have some differences in their correlation dynamics due to effect of external noise.
\subsection{GHZ-type initial states}
\label{ghz1}
In this section, we explore the effects of noise on the following three qubit GHZ-type initial states,
\begin{eqnarray}
\label{ghz2}
\hat{\rho}(0)=\frac{1-r}{8}\hat{I}_8+r\left|GHZ\right\rangle\left\langle GHZ\right|,
\end{eqnarray}
where $\left|GHZ\right\rangle=\frac{1}{\sqrt{2}}\left(\left|111\right\rangle+\left|000\right\rangle\right)$ is the GHZ-state, $r$ is the purity which ranges from 0 to 1 and $\hat{I}_8$ is the identity matrix of dimension $8$. For this initial state one can define the measurement operators as $\hat{M}_A\equiv\hat{\sigma}_y$ and $\hat{M}_A'\equiv\hat{\sigma}_x$~\cite{bb1}, then 
\begin{eqnarray}
\label{measurement1}
\hat{M}_A&=&\hat{\sigma}_y \otimes \hat{I}_2 \otimes \hat{I}_2,\nonumber\\
\hat{M}_A'&=&\hat{\sigma}_x \otimes \hat{I}_2 \otimes \hat{I}_2,\nonumber\\
\hat{M}_B&=&\hat{I}_2 \otimes [\cos(\theta_B)\hat{\sigma}_y-\sin(\theta_B)\hat{\sigma}_x] \otimes \hat{I}_2,\nonumber\\
\hat{M}_B'&=&\hat{I}_2 \otimes [\sin(\theta_B)\hat{\sigma}_y+\cos(\theta_B)\hat{\sigma}_x] \otimes \hat{I}_2,\nonumber\\
\hat{M}_C&=&\hat{I}_2 \otimes \hat{I}_2 \otimes [\cos(\theta_C)\hat{\sigma}_y-\sin(\theta_C)\hat{\sigma}_x],\nonumber\\
\hat{M}_C'&=&\hat{I}_2 \otimes \hat{I}_2 \otimes [\sin(\theta_C)\hat{\sigma}_y+\cos(\theta_C)\hat{\sigma}_x].
\end{eqnarray}
The expectation value of $\hat{B}$ and $\hat{S}$ operators for the GHZ-state under the classical noise can be easily obtained as:
\begin{eqnarray}
\label{exp1}
\left|\left\langle \hat{B}\right\rangle_{\rho}\right|&=&2re^{-3f(t)}\left|\cos(\theta_{BC})\right|,\nonumber\\
\left|\left\langle \hat{S}\right\rangle_{\rho}\right|&=&4re^{-3f(t)}\left|\cos(\theta_{BC})-\sin(\theta_{BC})\right|,
\end{eqnarray}
where $\theta_{BC}=\theta_B+\theta_C$. Also, it should be noted that this state does not have any bipartite correlations, however, it has non-zero tripartite negativity which is equal to
\begin{eqnarray}
\label{nega1}
N=\max\{0, -\frac{1}{8}(1-r-4re^{-3f(t)})\}.
\end{eqnarray}
\begin{figure}[!hbt]\centering
{\scalebox{0.8}{\includegraphics{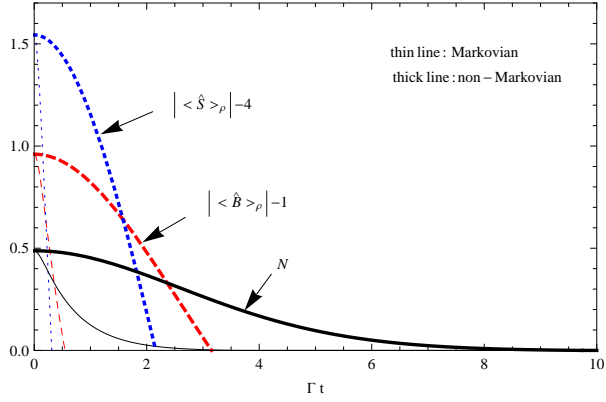}}}
\caption{The dynamics of $N$~(solid plots), $\left|\left\langle \hat{B}\right\rangle_{\rho}\right|-1$~(dashed plots) and $\left|\left\langle \hat{S}\right\rangle_{\rho}\right|-4$~(dotted plots) versus $\Gamma t$ with $r=0.98$ for GHZ-type initial state. Here the thick plots correspond to non-Markovian regime with $\gamma/\Gamma=0.1$ and the thin plots to Markovian regime with $\gamma/\Gamma=10$. }
\end{figure}
\begin{figure}[!hbt]\centering
{\scalebox{0.5}{\includegraphics{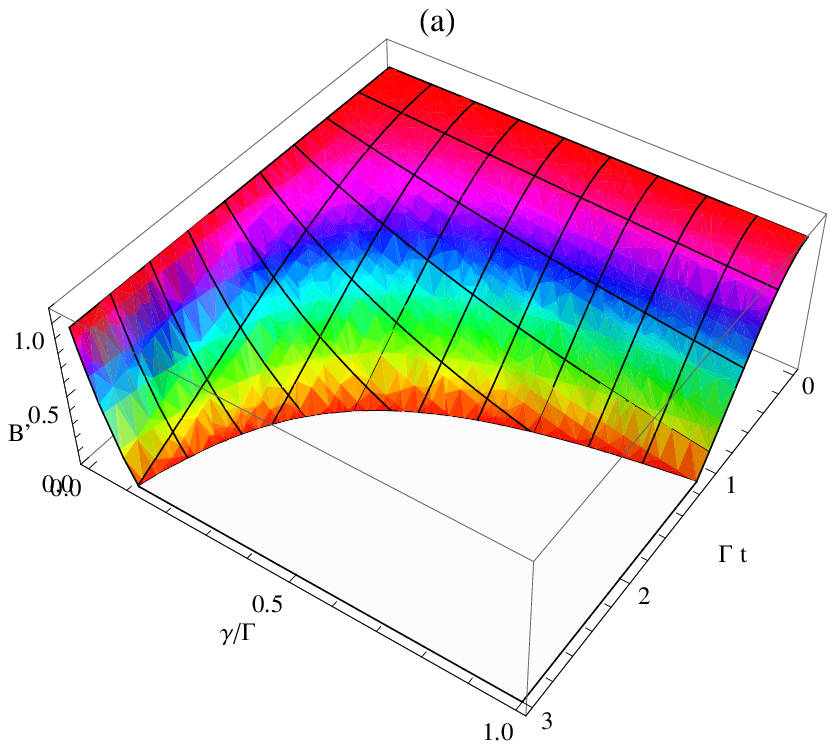}}}
{\scalebox{0.5}{\includegraphics{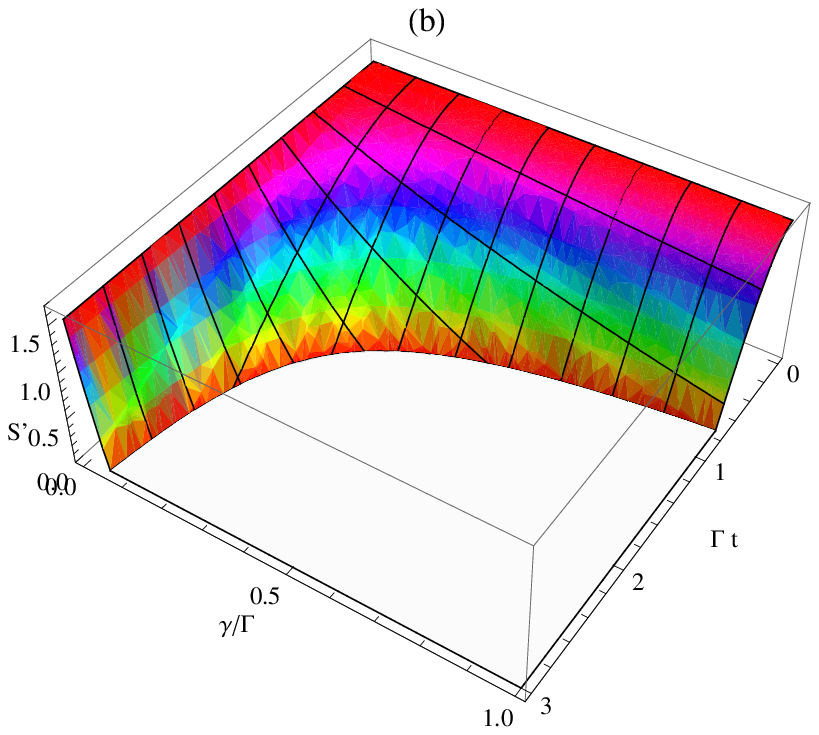}}}

{\scalebox{0.5}{\includegraphics{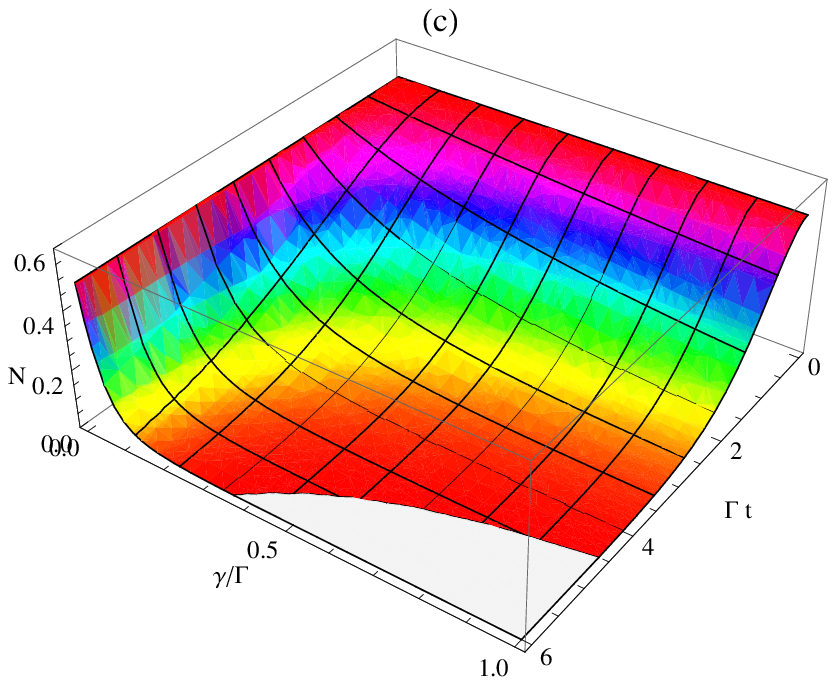}}}
\caption{The dynamics of $\left|\left\langle \hat{B}\right\rangle_{\rho}\right|-1$~(Fig.~(a)), $\left|\left\langle \hat{S}\right\rangle_{\rho}\right|-4$~(Fig.~(b)) and $N$~(Fig.~(c)) versus $\Gamma t$ and Markovianity, $\gamma/\Gamma$, with $r=0.98$ for GHZ-type initial state.}
\end{figure}

In Fig.~1 and~2, we have displayed the effects of non-Markovianity on the dynamics of tripartite entanglement and Bell nonlocalities for GHZ-type initial state with $r=0.98$. Both figures indicate that tripartite entanglement and Bell nonlocality violations suffer sudden death. It can be seen from Fig.~1 and~2 that the lifetime of tripartite entanglement as measured by the tripartite negativity is significantly longer than the lifetime of Bell nonlocalities in Markovian as well as non-Markovian coupling regimes. The non-Markovianity only delays the death, not prevent it, as can be seen from Fig.~2. 
\begin{figure}[!hbt]\centering
{\scalebox{0.5}{\includegraphics{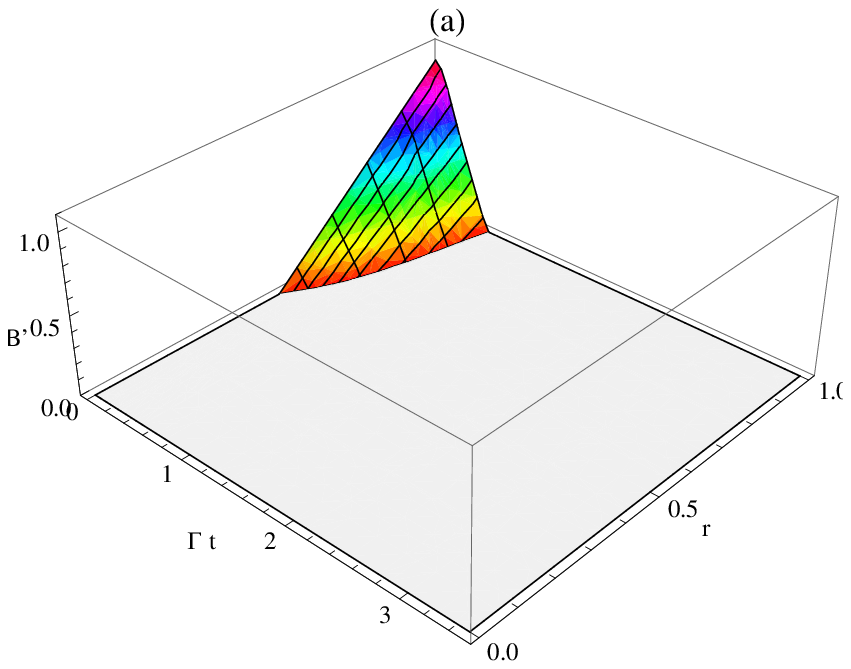}}}
{\scalebox{0.5}{\includegraphics{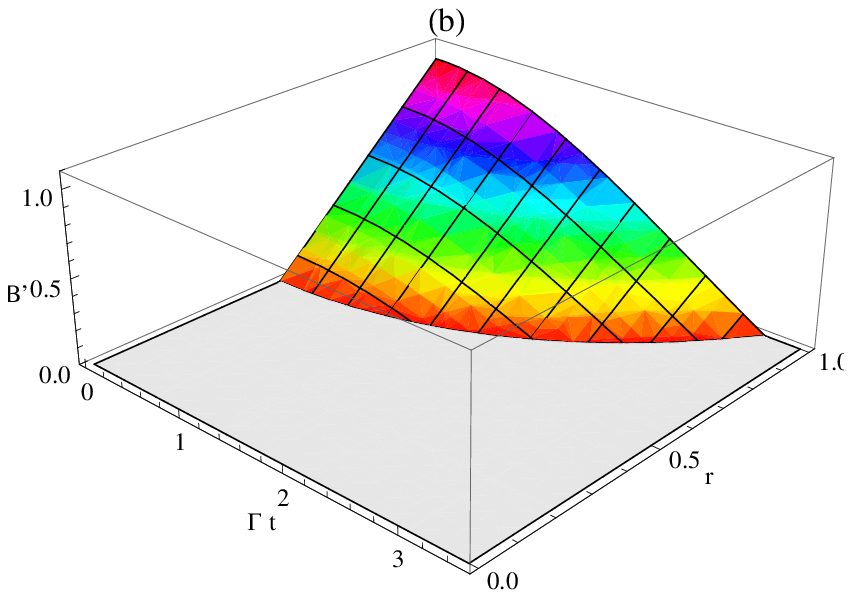}}}

{\scalebox{0.5}{\includegraphics{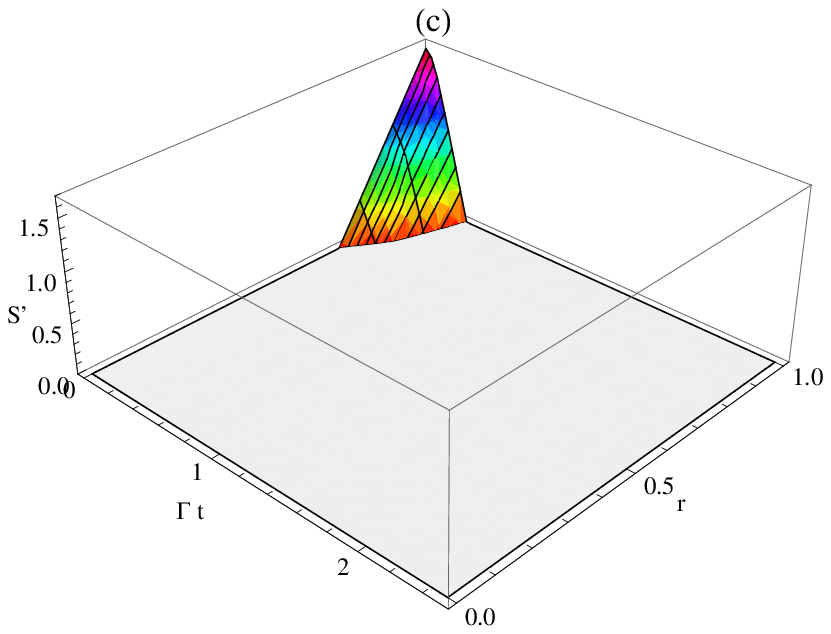}}}
{\scalebox{0.5}{\includegraphics{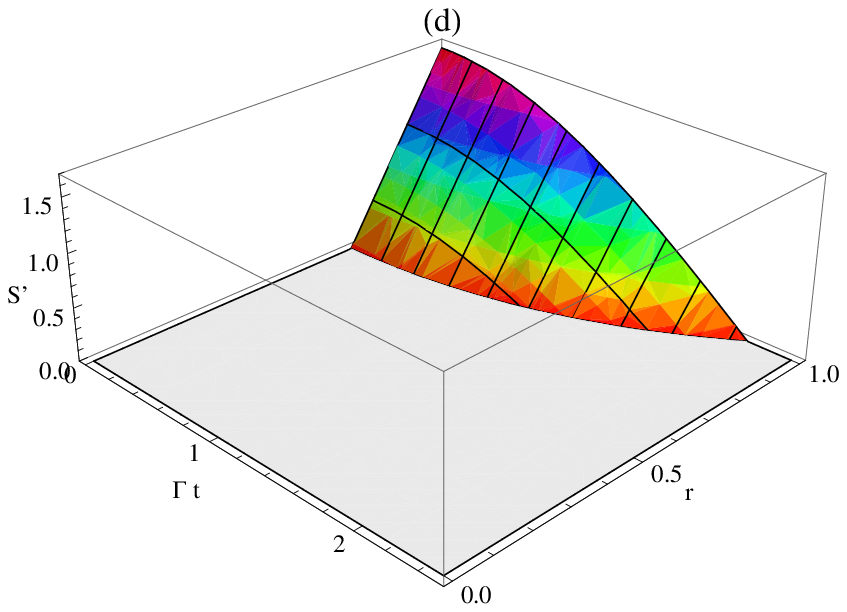}}}

{\scalebox{0.5}{\includegraphics{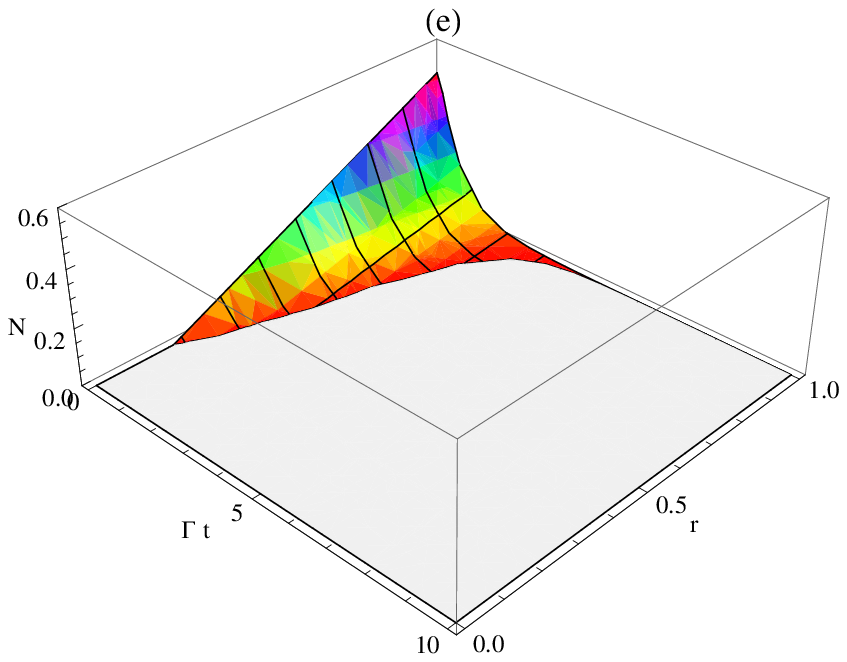}}}
{\scalebox{0.5}{\includegraphics{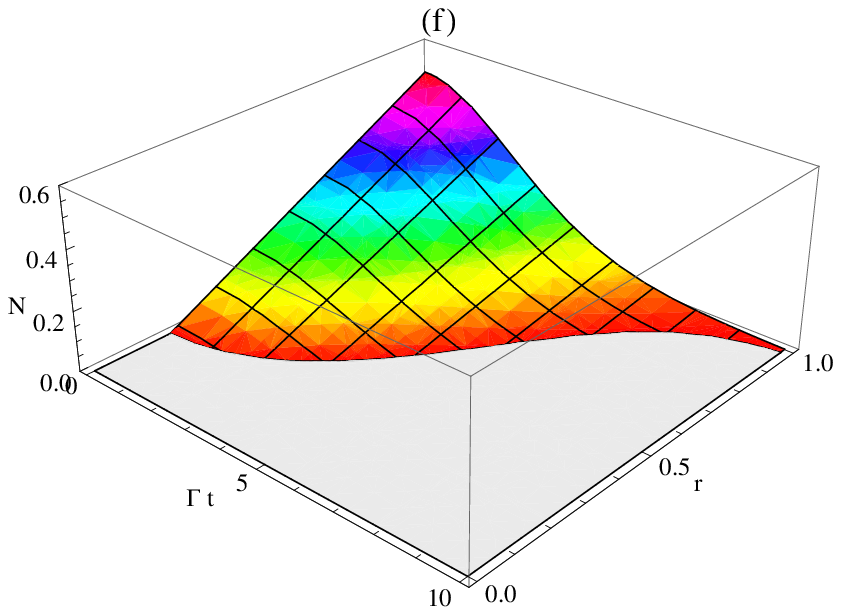}}}
\caption{The dynamics of $\left|\left\langle \hat{B}\right\rangle_{\rho}\right|-1$~(Fig.~(a) and~(b)), $\left|\left\langle \hat{S}\right\rangle_{\rho}\right|-4$~(Fig.~(c) and~(d)) and $N$~(Fig.~(e) and~(f)) versus $\Gamma t$ and $r$  for GHZ-type initial state. Fig.~(a),~(c) and~(e) correspond to Markovian regime with $\gamma/\Gamma=10$ and Fig.~(b),~(d) and~(f) to non-Markovian regime with $\gamma/\Gamma=0.1$.}
\end{figure}

The effect of purity is considered in Fig.~3, where we have plotted the tripartite negativity and Bell nonlocalities versus dimensionless time and the purity, $r$, for GHZ-type initial state in Markovian~($\gamma/\Gamma=10$) and non-Markovian~($\gamma/\Gamma=0.1$) coupling regimes. It can be seen that the purity is the most important parameter for the existence of Bell nonlocalities and the entanglement.  $\left|\left\langle \hat{B}\right\rangle_{\rho}\right|$ and $\left|\left\langle \hat{S}\right\rangle_{\rho}\right|$ suffer death at all time points for $r<0.5$ and $r<0.4$, respectively, while the non-zero values of tripartite entanglement exists for  a much wider range of purity values, $0.2\leq r \leq 1$. The range of $r$ for nonzero $N$ and Bell-inequality violations is independent of the Markovianity of the dynamics. Another important point is that for pure state~($r=1$), the Bell nonlocalities cease to exist in a finite time while entanglement dies only asymptotically.
\subsection{W-type initial states}
\label{w1}
The W-type initial state can be expressed as:
\begin{eqnarray}
\label{w}
\hat{\rho}(0)=\frac{1-r}{8}\hat{I}_8+r\left|W\right\rangle\left\langle W\right|,
\end{eqnarray}
where $\left|W\right\rangle=\frac{1}{\sqrt{3}}(\left|100\right\rangle+\left|010\right\rangle+\left|001\right\rangle)$ is the W-state and $r$ is the purity. For this type of initial states, the measurement operators for Bell nonlocalities are given by~\cite{bb2}
\begin{eqnarray}
\label{measurement2}
\hat{M}_A&=&\hat{\sigma}_z \otimes \hat{I}_2 \otimes \hat{I}_2,\nonumber\\
\hat{M}_A'&=&\hat{\sigma}_x \otimes \hat{I}_2 \otimes \hat{I}_2,\nonumber\\
\hat{M}_B&=&\hat{I}_2 \otimes [\cos(\theta_B)\hat{\sigma}_z-\sin(\theta_B)\hat{\sigma}_x] \otimes \hat{I}_2,\nonumber\\
\hat{M}_B'&=&\hat{I}_2 \otimes [\sin(\theta_B)\hat{\sigma}_z+\cos(\theta_B)\hat{\sigma}_x] \otimes \hat{I}_2,\nonumber\\
\hat{M}_C&=&\hat{I}_2 \otimes \hat{I}_2 \otimes [\cos(\theta_C)\hat{\sigma}_z-\sin(\theta_C)\hat{\sigma}_x],\nonumber\\
\hat{M}_C'&=&\hat{I}_2 \otimes \hat{I}_2 \otimes [\sin(\theta_C)\hat{\sigma}_z+\cos(\theta_C)\hat{\sigma}_x].
\end{eqnarray}
Then the time-dependent expectation values of the operators $\hat{B}$ and $\hat{S}$ can be calculated as:
\begin{eqnarray}
\label{exp2}
\left|\left\langle \hat{B}\right\rangle_{\rho}\right|&=&\frac{r}{2}(1+2e^{-2f(t)})\left|\sin(\theta_{BC})\right|,\nonumber\\
\left|\left\langle \hat{S}\right\rangle_{\rho}\right|&=&r(1+2e^{-2f(t)})\left|\left(\cos(\theta_{BC})+\sin(\theta_{BC})\right)\right|.
\end{eqnarray}

Unlike GHZ state, W state has a high degree of bipartite correlation. Thus, W-type states allow some comparison between tri- and bi-partite correlations. The analytic forms of the tri- and bi-partite entanglements and quantum discord for this state can be easily calculated as:
\begin{eqnarray}
\label{wnw}
N&=&\frac{1}{24}\max\{0,-3+3r+8\sqrt{2}re^{-2f(t)}\},\nonumber\\
C&=&\frac{1}{6}\max \{0,4re^{-2f(t)}-\sqrt{3(1-r)(3+r)} \},\nonumber\\
D&=&\min\{D_1,D_2\},
\end{eqnarray}
where
\begin{eqnarray}
\label{dp}
D_1&=&-\displaystyle\sum_{i=1}^2\lambda_i^A\log_2(\lambda_i^A)+\displaystyle\sum_{i=1}^4\lambda_i^{AB}\log_2(\lambda_i^{AB})-a\log_2\left(\frac{a}{a+b}\right)-b\log_2\left(\frac{b}{a+b}\right)\nonumber\\
&-&d\log_2\left(\frac{d}{b+d}\right)-b\log_2\left(\frac{b}{d+b}\right),\nonumber\\
D_2&=&-\displaystyle\sum_{i=1}^2\lambda_i^A\log_2(\lambda_i^A)+\displaystyle\sum_{i=1}^4\lambda_i^{AB}\log_2(\lambda_i^{AB})-\frac{1}{2}(1+\kappa)\log_2\left(\frac{1}{2}(1+\kappa)\right)\nonumber\\
&-&\frac{1}{2}(1-\kappa)\log_2\left(\frac{1}{2}(1-\kappa)\right),
\end{eqnarray}
where $\lambda_1^A=\frac{3-r}{6}, \lambda_2^A=\frac{3+r}{6}, \lambda_1^{AB}=\frac{1-r}{4}, \lambda_2^{AB}=\frac{3+r}{12}, \lambda_3^{AB}=\frac{1}{12}(3+r-4re^{-2f(t)}), \lambda_4^{AB}=\frac{1}{12}(3+r+4re^{-2f(t)}),a=\frac{1-r}{4},b=d=\frac{3+r}{12}, z=\frac{r}{3}e^{-2f(t)}$ and $\kappa=\frac{r}{3}\sqrt{1+4e^{-4f(t)}}$.
\begin{figure}[!hbt]\centering
{\scalebox{0.7}{\includegraphics{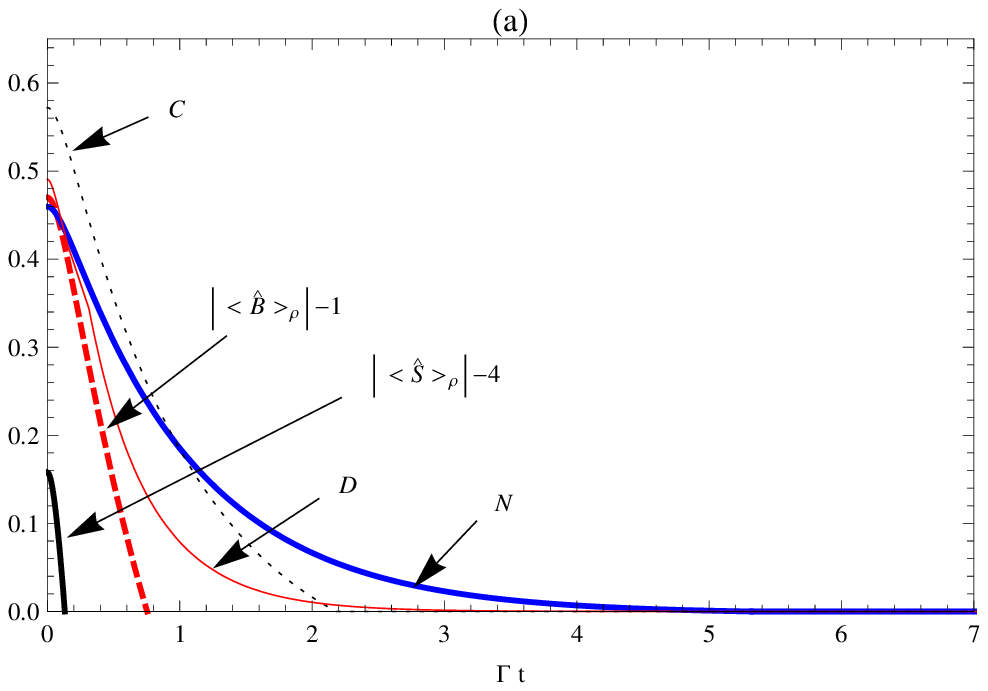}}}
{\scalebox{0.7}{\includegraphics{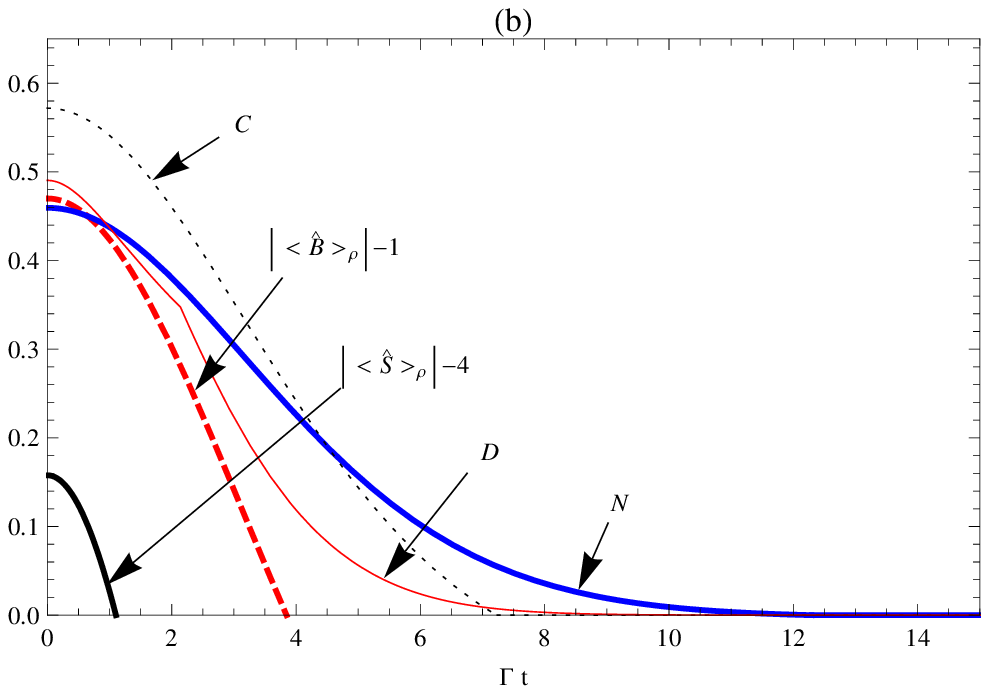}}}
\caption{(Color online) The dynamics of $N$~(thick blue solid plots), $C$~(thin black dotted plots), $D$~(thin red solid plots), $\left|\left\langle \hat{B}\right\rangle_{\rho}\right|-1$~(thick dashed red plots) and $\left|\left\langle \hat{S}\right\rangle_{\rho}\right|-4$~(thick black solid plots) versus $\Gamma t$ with $r=0.98$ for W-type initial state. Here Fig.~(a) corresponds to Markovian regime with $\gamma/\Gamma=10$ and Fig.~(b) to non-Markovian regime with $\gamma/\Gamma=0.1$.}
\end{figure}

The Markovian and non-Markovian dynamics of Bell nonlocalities, bi- and tri-partite entanglement and quantum discord for W-type initial state are displayed in Figs.~4(a) and~(b), respectively for $r=0.98$. One observation from these figures is the fact that tripartite entanglement as measured by  tripartite negativity has a longer lifetime compared to the bipartite entanglement as measured by concurrence, independent of Markovianity of the dynamics. One can also deduce from Figs.~4(a) and~(b) that both Bell nonlocalities  $\left|\left\langle \hat{B}\right\rangle_{\rho}\right|$ and $\left|\left\langle \hat{S}\right\rangle_{\rho}\right|$ are more fragile than all the other quantum correlations considered in the present study. Also, quantum discord is found to be immune to sudden death independent of the Markovianity of the dynamics. One should note that quantum discord has a discontinuity at $\Gamma t=0.4$ for Markovian and $\Gamma t=2.3$ for non-Markovian dynamics. The reason for the discontinuity is in the definition of quantum discord which involves $\min\{ D_1,D_2 \}$ where $D_1$ and $D_2$ are defined in the text. Ref.~\cite{expd} has shown experimentally that the discontinuity is real.
\begin{figure}[!hbt]\centering
{\scalebox{0.5}{\includegraphics{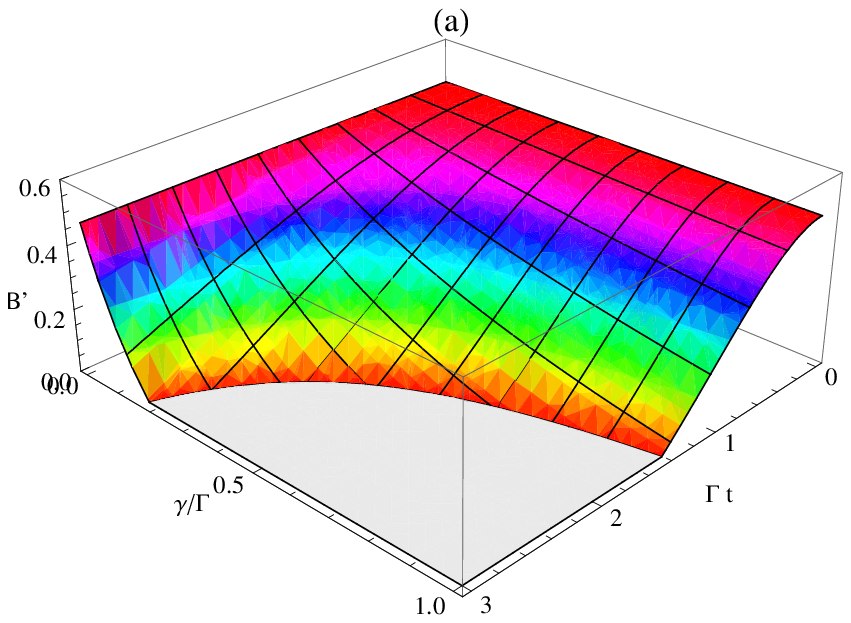}}}
{\scalebox{0.5}{\includegraphics{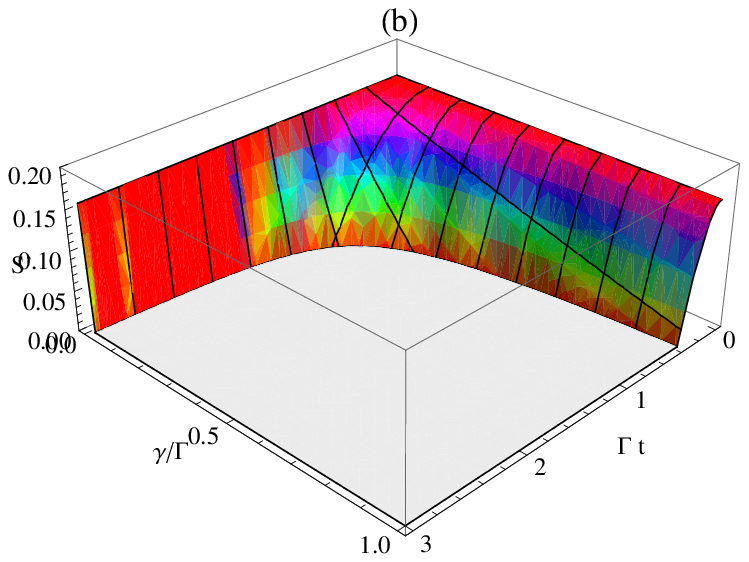}}}
\caption{The dynamics of $\left|\left\langle \hat{B}\right\rangle_{\rho}\right|-1$~(Fig.~(a)) and  $\left|\left\langle \hat{S}\right\rangle_{\rho}\right|-4$~(Fig.~(b))  versus $\Gamma t$ and $\gamma/\Gamma$ with $r=0.98$ for W-type initial state.}
\end{figure}
\begin{figure}[!hbt]\centering
{\scalebox{0.5}{\includegraphics{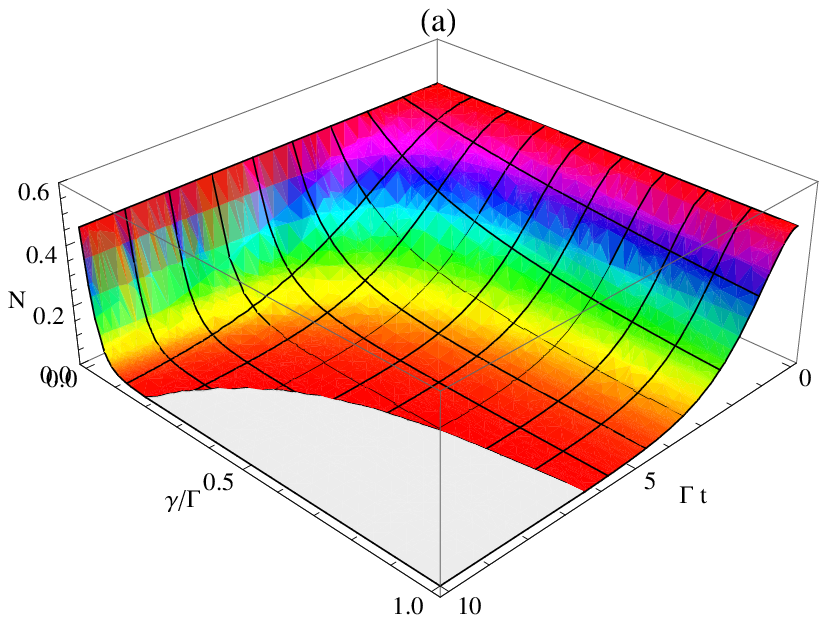}}}

{\scalebox{0.5}{\includegraphics{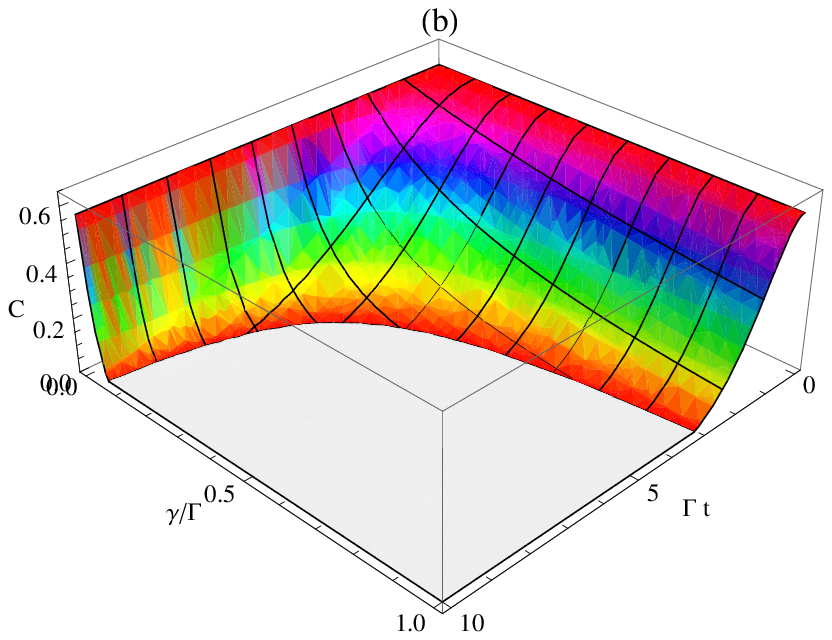}}}
{\scalebox{0.5}{\includegraphics{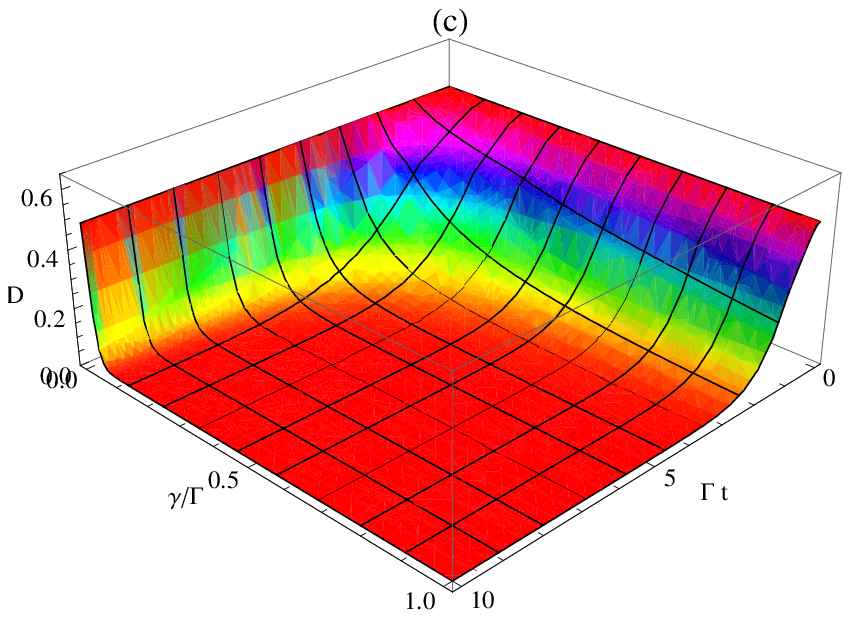}}}
\caption{The dynamics of $N$~(Fig.~(a)),  $C$~(Fig.~(b)) and $D$~(Fig.~(c))  versus $\Gamma t$ and $\gamma/\Gamma$ with $r=0.98$ for W-type initial state.}
\end{figure}

The effect of Markovianity on Bell nonlocalities and tripartite negativity, concurrence and quantum discord for W-type initial states are displayed as  functions of $\Gamma t$ and $\gamma/\Gamma$ in Fig.~5 and~6, respectively. The effect is found to be a prolongation of lifetime for the correlations that suffer sudden death~(Bell nonlocalities, $N$ and $C$).
\begin{figure}[!hbt]\centering
{\scalebox{0.5}{\includegraphics{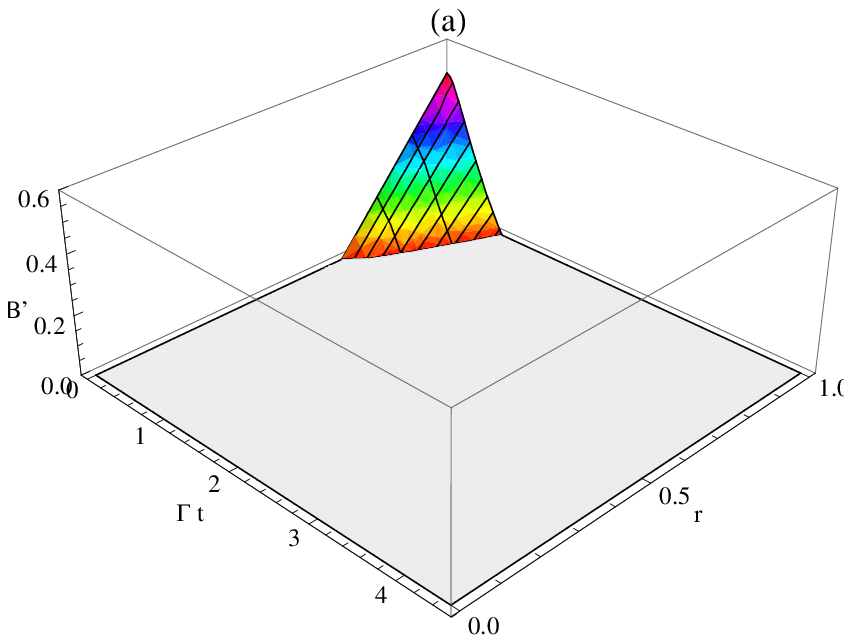}}}
{\scalebox{0.5}{\includegraphics{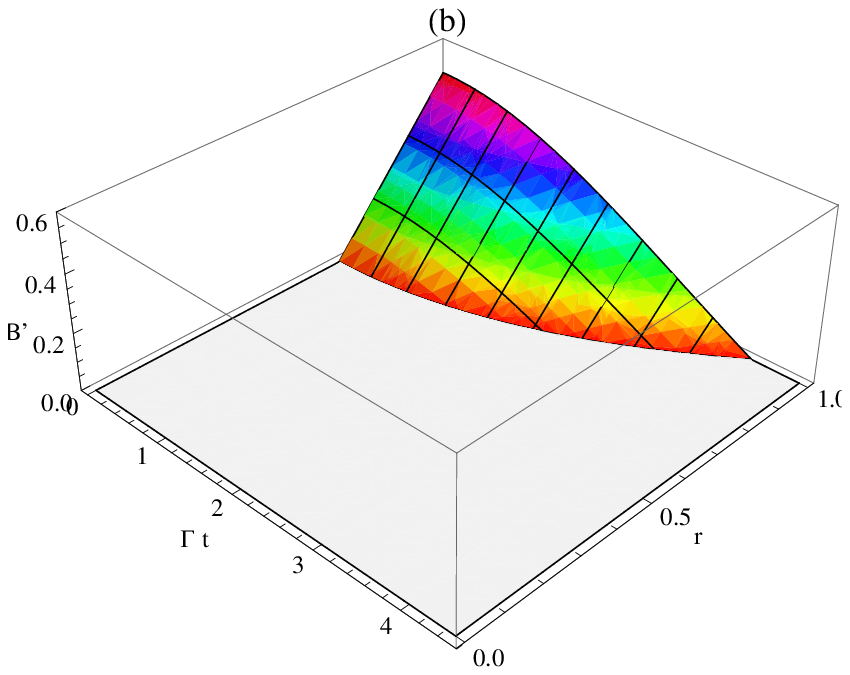}}}

{\scalebox{0.5}{\includegraphics{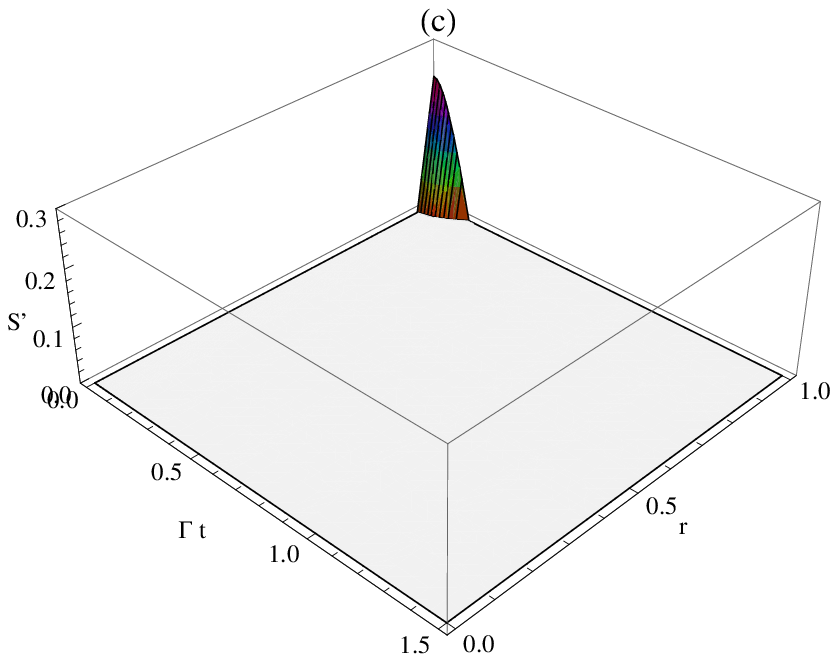}}}
{\scalebox{0.5}{\includegraphics{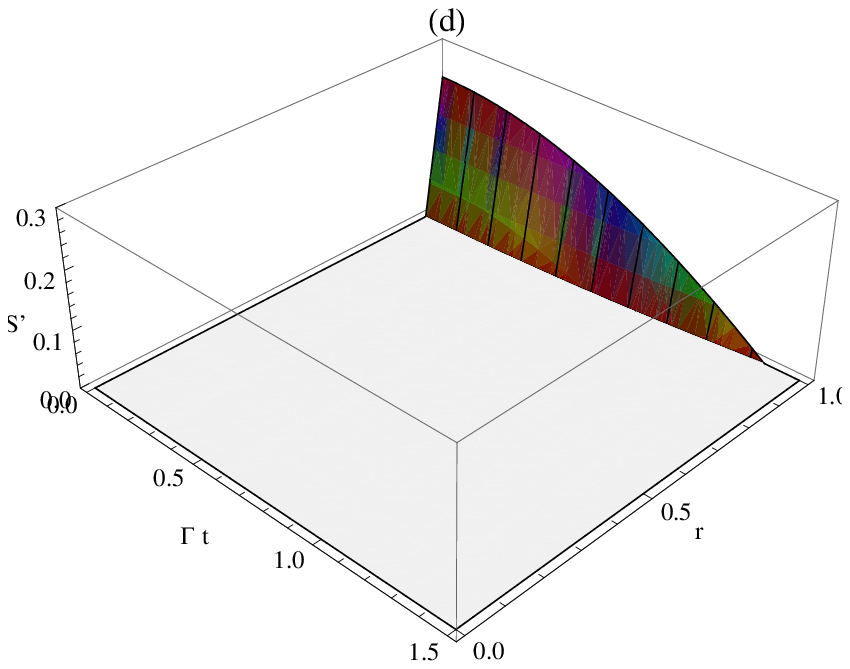}}}
\caption{The dynamics of $\left|\left\langle \hat{B}\right\rangle_{\rho}\right|-1$~(Fig.~(a) and~(b)) and $\left|\left\langle \hat{S}\right\rangle_{\rho}\right|-4$~(Fig.~(c) and~(d)) versus $\Gamma t$ and $r$  for W-type initial state. Fig.~(a) and~(c) correspond to Markovian regime with $\gamma/\Gamma=10$ and Fig.~(b) and~(d)  to non-Markovian regime with $\gamma/\Gamma=0.1$.}
\end{figure}

We have analyzed the purity dependence of Bell-nonlocalities as measured by $\left|\left\langle \hat{B}\right\rangle_{\rho}\right|$ and $\left|\left\langle \hat{S}\right\rangle_{\rho}\right|$ for the Markovian and non-Markovian dynamics and present the results in Fig.~7~(a)-(d). One can conclude two important findings from these figures: i) Both of nonlocalities are highly purity dependent; $\left|\left\langle \hat{B}\right\rangle_{\rho}\right|$ suffers death at all times in the regions where $r<0.7$, while the non-zero values of $\left|\left\langle \hat{S}\right\rangle_{\rho}\right|$ is limited to a much narrower range of $r$ values ($0.9<r\leq1$). These are independent of Markovianity of the dynamics. ii) As expected, non-Markovian dynamics offer a longer lived nonlocality compared to the Markovian case.
\begin{figure}[!hbt]\centering
{\scalebox{0.5}{\includegraphics{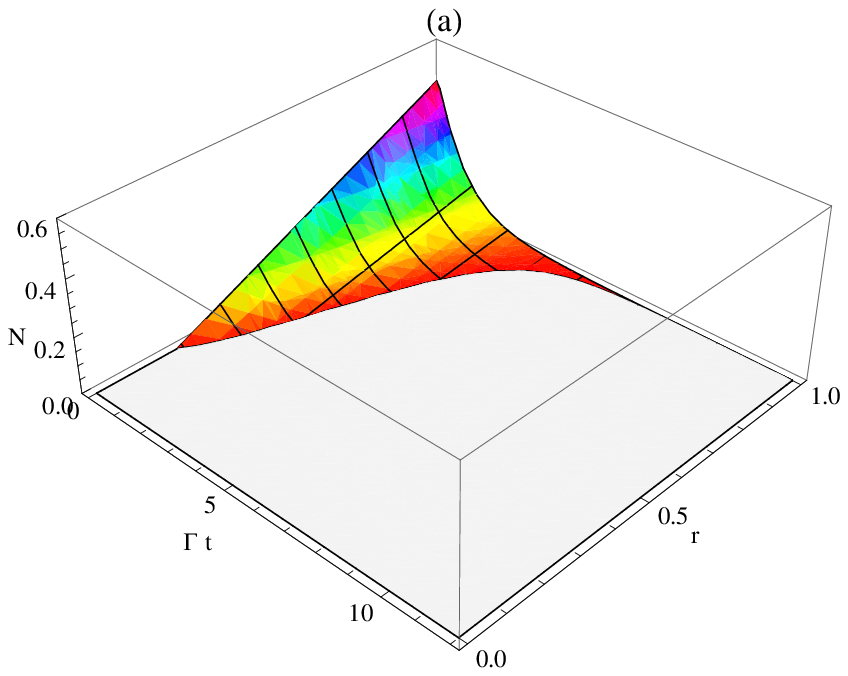}}}
{\scalebox{0.5}{\includegraphics{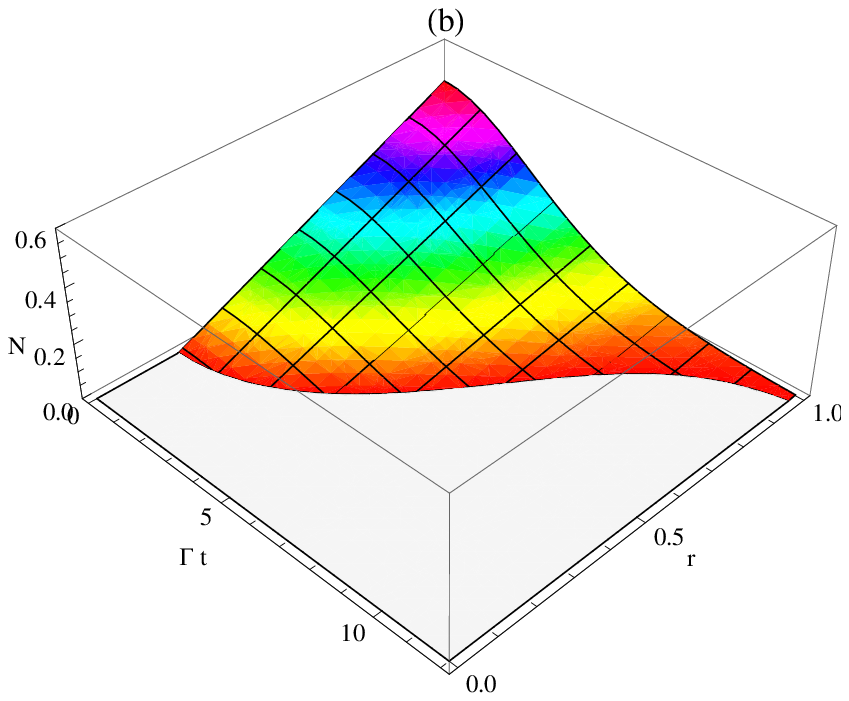}}}

{\scalebox{0.5}{\includegraphics{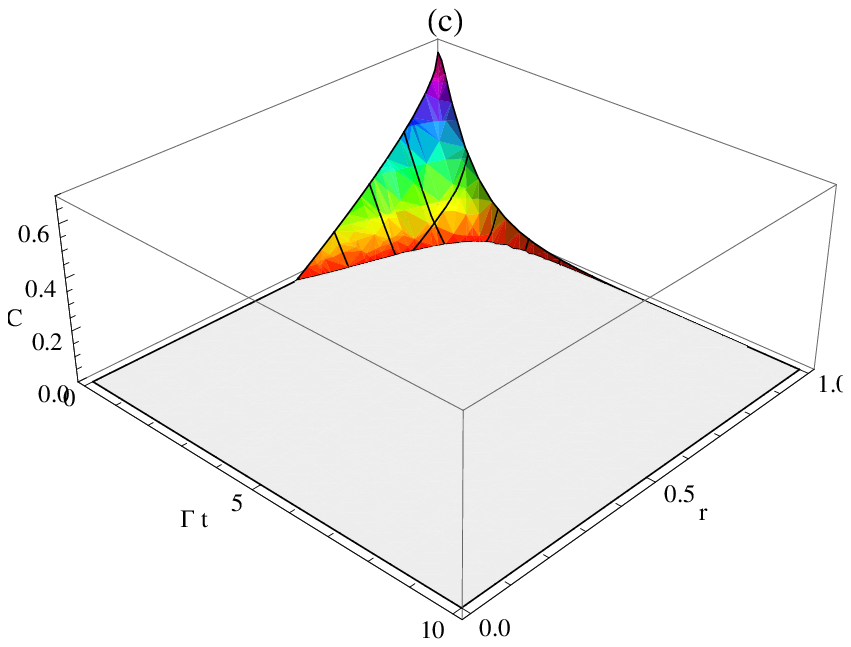}}}
{\scalebox{0.5}{\includegraphics{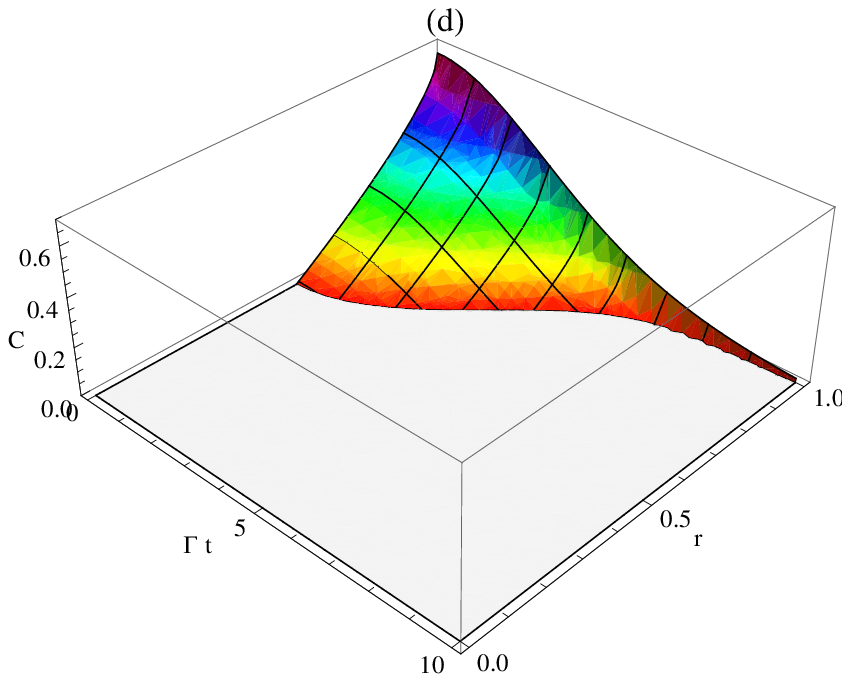}}}

{\scalebox{0.5}{\includegraphics{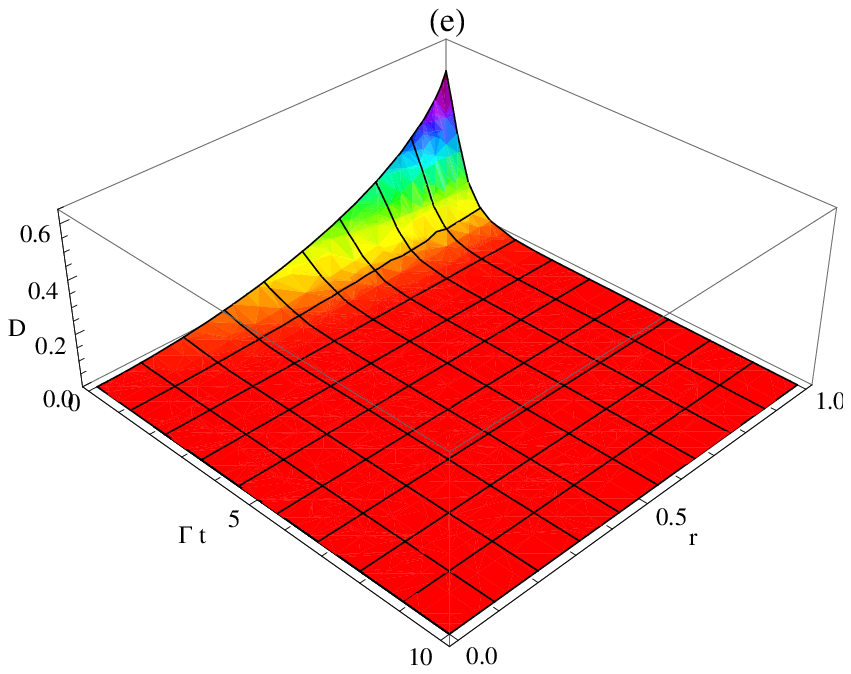}}}
{\scalebox{0.5}{\includegraphics{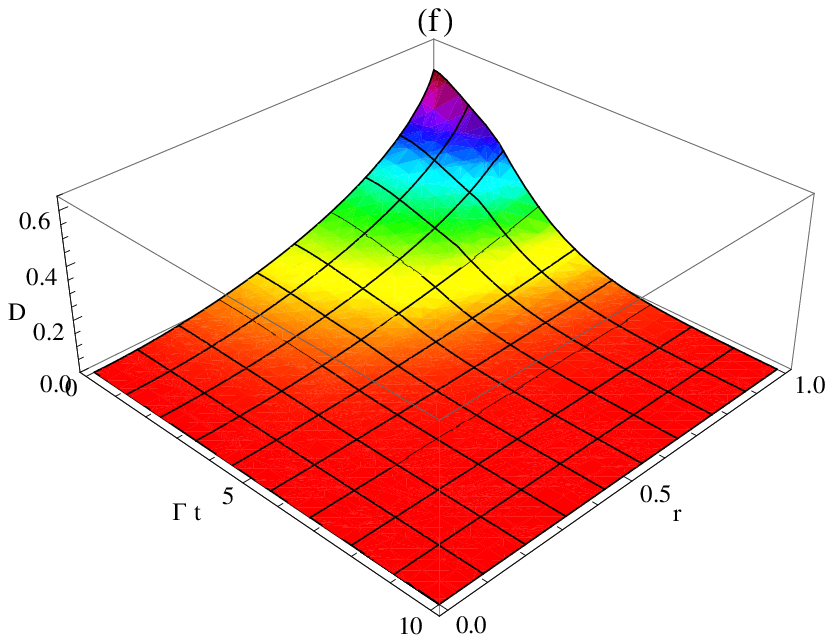}}}
\caption{The dynamics of $N$~(Fig.~(a) and~(b)), $C$~(Fig.~(c) and~(d)) and  $D$~(Fig.~(e) and~(f)) versus $\Gamma t$ and $r$  for W-type initial state. Fig.~(a),~(c) and~(e)  correspond to Markovian regime with $\gamma/\Gamma=10$ and Fig.~(b),~(d) and~(f)  to non-Markovian regime with $\gamma/\Gamma=0.1$. }
\end{figure}

Similar to the case of Bell-nonlocalities, we have considered the purity dependence of Markovian and non-Markovian dynamics of tripartite negativity, bipartite concurrence and quantum discord and displayed the results in Fig.8~(a)-(f). The most important finding from Fig.~8~(a)-(f) is the nonexistence of death for the quantum discord independent of purity of the initial states and Markovianity of the dynamics, which indicates that although entanglement type quantum correlations die out in finite time, not all types of quantum correlations are lost because of the noisy environment. Ferraro et al. showed that quantum discord can not have sudden death in Markovian environments~\cite{ferraro}. Our results indicate that absence of sudden death for quantum discord is valid for Markovian as well as non-Markovian dynamics for the particular states considered in the present work. Another important observation concerns the behavior of Bell-nonlocalities and entanglement measures for the pure initial states ($r=1$). While there is a sudden death of $\left|\left\langle \hat{B}\right\rangle_{\rho}\right|$ and $\left|\left\langle \hat{S}\right\rangle_{\rho}\right|$ for $r=1$, there is no ESD in $N$ or $C$. Also, nonzero bipartite entanglement is found for a narrower range of initial purity~($0.5\leq r \leq 1$) compared to nonzero tripartite entanglement~($0.2\leq r \leq 1$).

\section{Conclusions}
\label{conc}
We have analyzed the dynamics of quantum correlations, such as quantum discord, entanglement and Bell nonlocalities for three qubits that have stochastic time-dependent level spacings. The considered noise have Ornstein-Uhlenbeck type correlation. The dynamics is considered for GHZ- and W-type initial states and the survival times of the quantum correlations are compared. The tripartite entanglement is found to be immune to sudden death for pure GHZ as well as W states, while Bell inequalities cease to be violated for all types of initial states considered in this study. For GHZ-type initial states, there is no bipartite entanglement or quantum discord, while Bell nonlocality and tripartite entanglement as measured by tripartite negativity is nonzero for purity greater than approximately $0.4$ and $0.2$, respectively. Bell inequality is found to be not violated at all purities at much shorter times compared to the lifetime of entanglement.

W-type initial states display a richer dynamics, as they contain both bi- and tri-partite entanglement as well as Bell nonlocalities and nonzero quantum discord, initially. Quantum discord is observed to be more robust compared to entanglement, because it decays exponentially while the concurrence of the same state suffers sudden death. Also, tripartite entanglement is found to survive longer compared to the bipartite one if the state is mixed; for pure W states, both bi- and tri-partite entanglements decrease exponentially with time.

We have considered the effects of non-Markovian dynamics of both types of states and found that  the sole effect of non-Markovianity is to prolong the lifetime of quantum correlations compared to the Markovian dynamics, except for quantum discord which decreases only exponentially. Moreover, it should be noted that under Ornstein-Uhlenbeck type noise, once the entanglement and Bell nonlocalities die, rebirth or revival does not occur in their dynamics~\cite{noiseonly}.

\newpage

\appendix

\section{}
\label{procedure}
We consider a system that includes three subsystems $\widetilde{S}=\widetilde{A},\widetilde{B},\widetilde{C}$, and each subsystem includes one qubit, $S=A,B,C$, interacting with its local reservoir $R_s$ and there is no other interaction in the whole system. Initially, each qubit and its reservoir are independent, thus the evolution of the reduced density matrix for qubit $S$ is given by
 \begin{eqnarray}
\label{terdensity}
\hat{\rho}^S(t)=Tr_{R_S}\{\hat{U}^{\widetilde{S}}(t)\hat{\rho}^S(0)\otimes\hat{\rho}^{R_S}(0)\hat{U}^{\widetilde{S} \dagger}(t)\},
\end{eqnarray}
where the trace is taken over the reservoir $R_s$ degrees of freedom and $\hat{U}^{\widetilde{S}}(t)$ is the time evolution operator for the part $\widetilde{S}$. Eq.~(\ref{terdensity}) can be expressed in terms of the Kraus operator $\hat{K}_{\alpha \beta}^S(t)$ as
\begin{eqnarray}
\label{krausdensity}
\hat{\rho}^S(t)=\displaystyle\sum_{\alpha \beta} \hat{K}_{\alpha \beta}^S(t)\hat{\rho}^S(0)\hat{K}_{\alpha \beta}^{S\dagger }(t).
\end{eqnarray}
Since there is only interaction between single qubit and its corresponding reservoir, the time evolution operator $\hat{U}^T(t)$ of the complete system factorizes as 
\begin{eqnarray}
\hat{U}^T(t)=\hat{U}^{\widetilde{A}}(t)\otimes\hat{U}^{\widetilde{B}}(t)\otimes\hat{U}^{\widetilde{C}}(t),
\end{eqnarray}
then the reduced density matrix~(\ref{krausdensity}) for one qubit can be extended for three qubits as 
\begin{eqnarray}
\label{krausdensity3}
\hat{\rho}^T(t)=\displaystyle\sum_{\alpha_1 \beta_1}\sum_{\alpha_2 \beta_2} \sum_{\alpha_3 \beta_3}\hat{K}_{\alpha_1 \beta_1}^A(t)\hat{K}_{\alpha_2 \beta_2}^B(t)\hat{K}_{\alpha_3 \beta_3}^C(t)\hat{\rho}^T(0)\hat{K}_{\alpha_1 \beta_1}^{C\dagger}(t)\hat{K}_{\alpha_2 \beta_2}^{B\dagger}(t)\hat{K}_{\alpha_3 \beta_3}^{A\dagger}(t).
\end{eqnarray}
Given the basis $\{\left|1_S\right\rangle,\left|2_S\right\rangle\}$ for qubit $S$, inserting the identity operator $\hat{I}=\displaystyle\sum_i\left|i_S\right\rangle \left\langle i_S\right|$, one can get the reduced density matrix elements of the single qubit in the given basis as 
\begin{eqnarray}
\label{onedensity1}
\left\langle i_S\left|\hat{\rho}^{S}\right|i_S'\right\rangle=\rho^{S}_{i_Si_S'}(t)&=&\sum_{l_Sl_S'}A_{i_Si_S'}^{l_Sl_S'}(t)\rho^{S}_{l_Sl_S'}(0),
\end{eqnarray}
where $A_{i_Si_S'}^{l_Sl_S'}(t)=\displaystyle\sum_{\alpha_S\beta_S}\left\langle i_S\right|\hat{K}_{\alpha_S\beta_S}^S(t)\left|l_S\right\rangle \left\langle l_S'\right|\hat{K}_{\alpha_S\beta_S}^{S\dagger}(t)\left|i_S'\right\rangle$. Combining Eqs.~(\ref{krausdensity}),~(\ref{krausdensity3}) and ~(\ref{onedensity1}), we finally get the reduced density matrix elements of the three-qubit system
\begin{eqnarray}
\label{threedensity}
\left\langle i_1i_2i_3\left|\hat{\rho}(t)\right|i_1'i_2'i_3'\right\rangle&=&\nonumber\\
\rho_{i_1i_1',i_2i_2',i_3i_3'}(t)&=&\sum_{l_1l_1'}\sum_{l_2l_2'}\sum_{l_3l_3'} A_{i_1i_1'}^{l_1l_1'}(t)A_{i_2i_2'}^{l_2l_2'}(t)A_{i_3i_3'}^{l_3l_3'}(t)\rho_{l_1l_1',l_2l_2',l_3l_3'}(0).
\end{eqnarray}

The procedure given above allows us to obtain the dynamics of three qubits, provided that the dynamics of one qubit is known, by a purely algebraic way and independently from the initial conditions. The procedure for N-qudit system is given in Ref.~\cite{gprocedure}. 

We consider the single-qubit density matrix in the form,
\begin{eqnarray}\label{sqdms}
\hat{\rho}^S(t)=\left( \begin{array}{cc} \rho_{11}^S(t) & \rho_{12}^S(t) \\ \rho_{21}^S(t) & \rho_{22}^S(t) \end{array} \right),
\end{eqnarray}
where 
\begin{eqnarray}
\label{sqelements}
\rho_{11}^S(t)&=&u_t^S \rho_{11}^S(0)+v_t^S(t)\rho_{22}^S(0), \nonumber\\
\rho_{22}^S(t)&=&(1-u_t^S) \rho_{11}^S(0)+(1-v_t^S)\rho_{22}^S(0), \nonumber\\
\rho_{12}^S(t)&=&\rho_{21}^{S*}(t)=z_t^S\rho_{12}^S(0),
\end{eqnarray}
where $u_t^S,v_t^S$ and $z_t^S$ are functions of time and determined by the model of chosen.

By considering Eqs.~(\ref{onedensity1}),~(\ref{threedensity}) and~(\ref{sqelements}), in the standard basis $\left|1\right\rangle\equiv\left|111\right\rangle, \left|2\right\rangle\equiv\left|110\right\rangle, \left|3\right\rangle\equiv\left|101\right\rangle, \left|4\right\rangle\equiv\left|100\right\rangle, \left|5\right\rangle\equiv\left|011\right\rangle, \left|6\right\rangle\equiv\left|010\right\rangle, \left|7\right\rangle\equiv\left|001\right\rangle, \left|8\right\rangle\equiv\left|000\right\rangle $, we can obtain the diagonal elements of the reduced density matrix of three-qubit system as
\begin{eqnarray}
\label{tqdelements}
\rho_{11}(t)&=&u_t^Au_t^Bu_t^C\rho_{11}(0)+u_t^Au_t^Bv_t^C\rho_{22}(0)+u_t^Av_t^Bu_t^C\rho_{33}(0)+u_t^Av_t^Bv_t^C\rho_{44}(0)\nonumber\\ &+&v_t^Au_t^Bu_t^C\rho_{55}(0)+v_t^Au_t^Bv_t^C\rho_{66}(0)+v_t^Av_t^Bu_t^C\rho_{77}(0)+v_t^Av_t^Bv_t^C\rho_{88}(0),\nonumber\\
\rho_{22}(t)&=&u_t^Au_t^B(1-u_t^C)\rho_{11}(0)+u_t^Au_t^B(1-v_t^C)\rho_{22}(0)+u_t^Av_t^B(1-u_t^C)\rho_{33}(0)\nonumber\\&+&u_t^Av_t^B(1-v_t^C)\rho_{44}(0)+v_t^Au_t^B(1-u_t^C)\rho_{55}(0)+v_t^Au_t^B(1-v_t^C)\rho_{66}(0)\nonumber\\
&+&v_t^Av_t^B(1-u_t^C)\rho_{77}(0)+v_t^Av_t^B(1-v_t^C)\rho_{88}(0),\nonumber\\
\rho_{33}(t)&=&u_t^A(1-u_t^B)u_t^C\rho_{11}(0)+u_t^A(1-u_t^B)v_t^C\rho_{22}(0)+u_t^A(1-v_t^B)u_t^C\rho_{33}(0)\nonumber\\&+&u_t^A(1-v_t^B)v_t^C\rho_{44}(0)+v_t^A(1-u_t^B)u_t^C\rho_{55}(0)+v_t^A(1-u_t^B)v_t^C\rho_{66}(0)\nonumber\\
&+&v_t^A(1-v_t^B)u_t^C\rho_{77}(0)+v_t^A(1-v_t^B)v_t^C\rho_{88}(0),\nonumber\\
\rho_{44}(t)&=&u_t^A(1-u_t^B)(1-u_t^C)\rho_{11}(0)+u_t^A(1-u_t^B)(1-v_t^C)\rho_{22}(0)\nonumber\\
&+&u_t^A(1-v_t^B)(1-u_t^C)\rho_{33}(0)+u_t^A(1-v_t^B)(1-v_t^C)\rho_{44}(0)\nonumber\\
&+&v_t^A(1-u_t^B)(1-u_t^C)\rho_{55}(0)+v_t^A(1-u_t^B)(1-v_t^C)\rho_{66}(0)\nonumber\\
&+&v_t^A(1-v_t^B)(1-u_t^C)\rho_{77}(0)+v_t^A(1-v_t^B)(1-v_t^C)\rho_{88}(0),\nonumber\\
\rho_{55}(t)&=&(1-u_t^A)u_t^Bu_t^C\rho_{11}(0)+(1-u_t^A)u_t^Bv_t^C\rho_{22}(0)+(1-u_t^A)v_t^Bu_t^C\rho_{33}(0)\nonumber\\
&+&(1-u_t^A)v_t^Bv_t^C\rho_{44}(0)+(1-v_t^A)u_t^Bu_t^C\rho_{55}(0)+(1-v_t^A)u_t^Bv_t^C\rho_{66}(0)\nonumber\\
&+&(1-v_t^A)v_t^Bu_t^C\rho_{77}(0)+(1-v_t^A)v_t^Bv_t^C\rho_{88}(0),\nonumber\\
\rho_{66}(t)&=&(1-u_t^A)u_t^B(1-u_t^C)\rho_{11}(0)+(1-u_t^A)u_t^B(1-v_t^C)\rho_{22}(0)\nonumber\\
&+&(1-u_t^A)v_t^B(1-u_t^C)\rho_{33}(0)+(1-u_t^A)v_t^B(1-v_t^C)\rho_{44}(0)\nonumber\\
&+&(1-v_t^A)u_t^B(1-u_t^C)\rho_{55}(0)+(1-v_t^A)u_t^B(1-v_t^C)\rho_{66}(0)\nonumber\\
&+&(1-v_t^A)v_t^B(1-u_t^C)\rho_{77}(0)+(1-v_t^A)v_t^B(1-v_t^C)\rho_{88}(0),\nonumber\\
\rho_{77}(t)&=&(1-u_t^A)(1-u_t^B)u_t^C\rho_{11}(0)+(1-u_t^A)(1-u_t^B)v_t^C\rho_{22}(0)\nonumber\\
&+&(1-u_t^A)(1-v_t^B)u_t^C\rho_{33}(0)+(1-u_t^A)(1-v_t^B)v_t^C\rho_{44}(0)\nonumber\\
&+&(1-v_t^A)(1-u_t^B)u_t^C\rho_{55}(0)+(1-v_t^A)(1-u_t^B)v_t^C\rho_{66}(0)\nonumber\\
&+&(1-v_t^A)(1-v_t^B)u_t^C\rho_{77}(0)+(1-v_t^A)(1-v_t^B)v_t^C\rho_{88}(0),\nonumber\\
\rho_{88}(t)&=&(1-u_t^A)(1-u_t^B)(1-u_t^C)\rho_{11}(0)+(1-u_t^A)(1-u_t^B)(1-v_t^C)\rho_{22}(0)\nonumber\\
&+&(1-u_t^A)(1-v_t^B)(1-u_t^C)\rho_{33}(0)+(1-u_t^A)(1-v_t^B)(1-v_t^C)\rho_{44}(0)\nonumber\\
&+&(1-v_t^A)(1-u_t^B)(1-u_t^C)\rho_{55}(0)+(1-v_t^A)(1-u_t^B)(1-v_t^C)\rho_{66}(0)\nonumber\\
&+&(1-v_t^A)(1-v_t^B)(1-u_t^C)\rho_{77}(0)+(1-v_t^A)(1-v_t^B)(1-v_t^C)\rho_{88}(0),\nonumber\\
\end{eqnarray}
and the off-diagonal elements as
\begin{eqnarray}
\label{tqoelements}
\rho_{12}(t)&=&u_t^Au_t^Bz_t^C\rho_{12}(0)+u_t^Av_t^Bz_t^C\rho_{34}(0)+v_t^Au_t^Bz_t^C\rho_{56}(0)+v_t^Av_t^Bz_t^C\rho_{78}(0),\nonumber\\
\rho_{13}(t)&=&u_t^Az_t^Bu_t^C\rho_{13}(0)+u_t^Az_t^Bv_t^C\rho_{24}(0)+v_t^Az_t^Bu_t^C\rho_{57}(0)+v_t^Az_t^Bv_t^C\rho_{68}(0),\nonumber\\
\rho_{14}(t)&=&u_t^Az_t^Bz_t^C\rho_{14}(0)+v_t^Az_t^Bz_t^C\rho_{58}(0),\nonumber\\
\rho_{15}(t)&=&z_t^Au_t^Bu_t^C\rho_{15}(0)+z_t^Au_t^Bv_t^C\rho_{26}(0)+z_t^Av_t^Bu_t^C\rho_{37}(0)+z_t^Av_t^Bv_t^C\rho_{48}(0),\nonumber\\
\rho_{16}(t)&=&z_t^Au_t^Bz_t^C\rho_{16}(0)+z_t^Av_t^Bz_t^C\rho_{38}(0),\nonumber\\
\rho_{17}(t)&=&z_t^Az_t^Bu_t^C\rho_{17}(0)+z_t^Az_t^Bv_t^C\rho_{28}(0),\nonumber\\
\rho_{18}(t)&=&z_t^Az_t^Bz_t^C\rho_{18}(0),\nonumber\\
\rho_{23}(t)&=&u_t^Az_t^Bz_t^{C*}\rho_{23}(0)+v_t^Az_t^Bz_t^{C*}\rho_{67}(0),\nonumber\\
\rho_{24}(t)&=&u_t^Az_t^B(1-u_t^C)\rho_{13}(0)+u_t^Az_t^B(1-v_t^C)\rho_{24}(0)+v_t^Az_t^B(1-u_t^C)\rho_{57}(0)\nonumber\\
&+&v_t^Az_t^B(1-v_t^C)\rho_{68}(0),\nonumber\\
\rho_{25}(t)&=&z_t^Au_t^Bz_t^{C*}\rho_{25}(0)+z_t^Av_t^Bz_t^{C*}\rho_{47}(0),\nonumber\\
\rho_{26}(t)&=&z_t^Au_t^B(1-u_t^C)\rho_{15}(0)+z_t^Au_t^B(1-v_t^C)\rho_{26}(0)+z_t^Av_t^B(1-u_t^C)\rho_{37}(0)\nonumber\\
&+&z_t^Av_t^B(1-v_t^C)\rho_{48}(0),\nonumber\\
\rho_{27}(t)&=&z_t^Az_t^Bz_t^{C*}\rho_{27}(0),\nonumber\\
\rho_{28}(t)&=&z_t^Az_t^B(1-u_t^C)\rho_{17}(0)+z_t^Az_t^B(1-v_t^C)\rho_{28}(0),\nonumber\\
\rho_{34}(t)&=&u_t^A(1-u_t^B)z_t^C\rho_{12}(0)+u_t^A(1-v_t^B)z_t^C\rho_{34}(0)+v_t^A(1-u_t^B)z_t^C\rho_{56}(0)\nonumber\\
&+&v_t^A(1-v_t^B)z_t^C\rho_{78}(0),\nonumber\\
\rho_{35}(t)&=&z_t^Az_t^{B*}u_t^C\rho_{35}(0)+z_t^Az_t^{B*}v_t^C\rho_{46}(0),\nonumber\\
\rho_{36}(t)&=&z_t^Az_t^{B*}z_t^C\rho_{36}(0),\nonumber\\
\rho_{37}(t)&=&z_t^A(1-u_t^B)u_t^C\rho_{15}(0)+z_t^A(1-u_t^B)v_t^C\rho_{26}(0)+z_t^A(1-v_t^B)u_t^C\rho_{37}(0)\nonumber\\
&+&z_t^A(1-v_t^B)v_t^C\rho_{48}(0),\nonumber\\
\rho_{38}(t)&=&z_t^A(1-u_t^B)z_t^C\rho_{16}(0)+z_t^A(1-v_t^B)z_t^C\rho_{38}(0),\nonumber\\
\rho_{45}(t)&=&z_t^Az_t^{B*}z_t^{C*}\rho_{45}(0),\nonumber\\
\rho_{46}(t)&=&z_t^Az_t^{B*}(1-u_t^C)\rho_{35}(0)+z_t^Az_t^{B*}(1-v_t^C)\rho_{46}(0),\nonumber\\
\rho_{47}(t)&=&z_t^A(1-u_t^B)z_t^{C*}\rho_{25}(0)+z_t^A(1-v_t^B)z_t^{C*}\rho_{47}(0),\nonumber\\
\rho_{48}(t)&=&z_t^A(1-u_t^B)(1-u_t^C)\rho_{15}(0)+z_t^A(1-u_t^B)(1-v_t^C)\rho_{26}(0)\nonumber\\
&+&z_t^A(1-v_t^B)(1-u_t^C)\rho_{37}(0)+z_t^A(1-v_t^B)(1-v_t^C)\rho_{48}(0),\nonumber\\
\rho_{56}(t)&=&(1-u_t^A)u_t^Bz_t^C\rho_{12}(0)+(1-u_t^A)v_t^Bz_t^C\rho_{34}(0)+(1-v_t^A)u_t^Bz_t^C\rho_{56}(0)\nonumber\\
&+&(1-v_t^A)v_t^Bz_t^C\rho_{78}(0),\nonumber\\
\rho_{57}(t)&=&(1-u_t^A)z_t^Bu_t^C\rho_{13}(0)+(1-u_t^A)z_t^Bv_t^C\rho_{24}(0)+(1-v_t^A)z_t^Bu_t^C\rho_{57}(0)\nonumber\\
&+&(1-v_t^A)z_t^Bv_t^C\rho_{68}(0),\nonumber\\
\rho_{58}(t)&=&(1-u_t^A)z_t^Bz_t^C\rho_{14}(0)+(1-v_t^A)z_t^Bz_t^C\rho_{58}(0),\nonumber\\
\rho_{67}(t)&=&(1-u_t^A)z_t^Bz_t^{C*}\rho_{23}(0)+(1-v_t^A)z_t^Bz_t^{C*}\rho_{67}(0),\nonumber\\
\rho_{68}(t)&=&(1-u_t^A)z_t^B(1-u_t^C)\rho_{13}(0)+(1-u_t^A)z_t^B(1-v_t^C)\rho_{24}(0)\nonumber\\
&+&(1-v_t^A)z_t^B(1-u_t^C)\rho_{57}(0)+(1-v_t^A)z_t^B(1-v_t^C)\rho_{68}(0),\nonumber\\
\rho_{78}(t)&=&(1-u_t^A)(1-u_t^B)z_t^C\rho_{12}(0)+(1-u_t^A)(1-v_t^B)z_t^C\rho_{34}(0)\nonumber\\
&+&(1-v_t^A)(1-u_t^B)z_t^C\rho_{56}(0)+(1-v_t^A)(1-v_t^B)z_t^C\rho_{78}(0).
\end{eqnarray}

\end{document}